\documentclass[sigconf]{acmart}
\copyrightyear{2022}
\acmYear{2022}
\setcopyright{acmcopyright}
\acmConference[CCS '22] {Proceedings of the 2022 ACM SIGSAC Conference on Computer and Communications Security}{November 7--11, 2022}{Los Angeles, CA, USA.}
\acmBooktitle{Proceedings of the 2022 ACM SIGSAC Conference on Computer and Communications Security (CCS '22), November 7--11, 2022, Los Angeles, CA, USA}
\acmPrice{15.00}
\acmISBN{978-1-4503-9450-5/22/11}
\acmDOI{10.1145/3548606.3560613}

\usepackage[linesnumbered,ruled,vlined]{algorithm2e}
\usepackage{color, colortbl}
\usepackage{amsmath,amsfonts}
\usepackage{mathtools}
\usepackage{balance}
\usepackage{caption}
\usepackage{listings}
\usepackage{subcaption}
\usepackage{graphicx}
\usepackage{xspace}
\usepackage{booktabs}
\usepackage{multirow}
\usepackage{xcolor}
\newcommand*\circles[1]{\textcircled{#1}}

\newcommand{\extag}[1]{\texttt{tag}(#1)}
\newcommand{\expage}[1]{\texttt{page}(#1)}
\newcommand{\tageq}[1]{\texttt{sameTag}(#1)}
\newcommand{\seteq}[1]{\texttt{sameSet}(#1)}
\newcommand{\pageeq}[1]{\texttt{samePage}(#1)}
\newcommand{\exbus}[1]{\texttt{bus}(#1)}
\newcommand{\exword}[1]{\texttt{word}(#1)}
\newcommand{\exidx}[1]{\texttt{set}(#1)}

\newcommand{\tool}{Scam-V}
\newcommand{\framework}{\textsc{Plumber}\xspace}

\newcommand{\request}{Generative Testcase Specification}

\newcommand{\requestShort}{GTS}

\newcommand{\new}[1]{#1}

\newcommand{\regname}[1]{\lstinline{#1}}

\newcommand{\pluseq}{\mathrel{+}=}

\newcommand{\parhead}[1]  {\textbf{#1}}

\let\orgautoref\autoref
\renewcommand{\autoref}
        {\def\equationautorefname{Equation}%
         \def\figureautorefname{Fig.}%
         \def\subfigureautorefname{Fig.}%
         \def\Itemautorefname{Item}%
         \def\tableautorefname{Table}%
         \def\algorithmautorefname{Algorithm}%
         \def\paragraphautorefname{Paragraph}%
         \def\sectionautorefname{\S}%
         \def\subsectionautorefname{\S}%
         \def\subsubsectionautorefname{\S}%
         \def\chapterautorefname{Chapter}%
         \def\partautorefname{Part}%
         \def\goalautorefname{Goal}%
         \def\reqautorefname{Req.}%
         \def\adviceautorefname{Rule}%
         \def\parameterautorefname{Param.}%
         \def\definitionautorefname{Definition}%
         \def\definitionsautorefname{Definition}%
         \def\propertyautorefname{Property}%
         \orgautoref}
\newcommand{\styleSet}[1]        {\ensuremath{\mathcal{#1}}}
\newcommand{\styleEntity}[1]     {\ensuremath{\mathit{#1}}}
\newcommand{\styleAlgorithm}[1]  {\ensuremath{\mathsf{#1}}}

\newcommand{\styleLogic}[1] {\ensuremath{\texttt{#1}}}

\newcommand{\nocc}              {\styleAlgorithm{occ}\xspace}
\newcommand{\npos}              {\styleAlgorithm{n_{pos}}\xspace}
\newcommand{\nppos}              {\styleAlgorithm{n'_{pos}}\xspace}
\newcommand{\total}              {\styleAlgorithm{Total}\xspace}

\newcommand{\bits}[3]             {\ensuremath{{#3}^{{#1}-{#2}}}}

\newcommand{\prog}              {\styleSet{P}\xspace}
\newcommand{\proghat}              {\styleSet{P}(A)\xspace}
\newcommand{\behavior}              {\styleSet{B}\xspace}
\newcommand{\relation}              {\styleSet{R}(A, b)\xspace}
\newcommand{\inputzero}              {\styleSet{I}}

\newcommand{\ltag}              {\styleAlgorithm{Tag}}
\newcommand{\lset}              {\styleAlgorithm{Set}}
\newcommand{\ltags}              {\styleAlgorithm{Tags}}
\newcommand{\lsets}              {\styleAlgorithm{Sets}}

\newcommand{\ce}              {\styleAlgorithm{PR}\xspace}
\newcommand{\nce}              {\styleAlgorithm{nPR}\xspace}
\newcommand{\addr}[1]              {\ensuremath{\styleEntity{a_{#1}}}\xspace}

\newcommand{\NOP}              {\lstinline|nop|\xspace}

\newcommand{\LT}{LT}
\newcommand{\counterexample}{leakage witness}

\newcommand{\dir}             {\ensuremath{\styleLogic{dir}}\xspace}
\newcommand{\dirone}[1]             {\ensuremath{\styleLogic{dir}_{#1}}}

\newcommand{\attr}              {\styleLogic{attr}}

\newcommand{\branchthree}[3]             {\ensuremath{\styleLogic{B}_{#1,#2,#3}}}
\newcommand{\setbranchtwo}[2]             {\ensuremath{\styleLogic{S}_{#1,#2}}}
\newcommand{\bidone}              {\styleLogic{c\textsubscript{1}}}
\newcommand{\bidtwo}              {\styleLogic{c\textsubscript{2}}}

\newcommand{\bool}              {\styleLogic{bool}}
\newcommand{\bfalse}              {\styleLogic{F}}
\newcommand{\btrue}              {\styleLogic{T}}
\newcommand{\bStep}              {\styleLogic{step}}
\newcommand{\mem}             {\ensuremath{\styleLogic{M}}\xspace}
\newcommand{\memone}[1]             {\ensuremath{\styleLogic{M}_{#1}}}
\newcommand{\memtwo}[2]             {\ensuremath{\styleLogic{M}_{#1,#2}}}
\newcommand{\arith}              {\styleLogic{A}\xspace}\newcommand{\arithtwo}[2]             {\ensuremath{\styleLogic{A}_{#1,#2}}}
\newcommand{\nop}              {\styleLogic{N}\xspace}

\newcommand{\mytag}              {\styleLogic{t}}
\newcommand{\tagone}              {\styleLogic{t\textsubscript{1}}}
\newcommand{\tagtwo}              {\styleLogic{t\textsubscript{2}}}

\newcommand{\valueone}              {\styleLogic{v\textsubscript{1}}}
\newcommand{\valuetwo}              {\styleLogic{v\textsubscript{2}}}
\newcommand{\tagthree}              {\styleLogic{t\textsubscript{3}}}
\newcommand{\tagfour}              {\styleLogic{t\textsubscript{4}}}
\newcommand{\tagfive}              {\styleLogic{t\textsubscript{5}}}
\newcommand{\tagsix}              {\styleLogic{t\textsubscript{6}}}
\newcommand{\tagseven}              {\styleLogic{t\textsubscript{7}}}

\newcommand{\tagm}              {\styleLogic{t\textsubscript{m}}}
\newcommand{\tagx}              {\styleLogic{t\textsubscript{x}}}

\newcommand{\tagxb}              {\styleLogic{t\textsubscript{x+b}}}

\newcommand{\setone}              {\styleLogic{s\textsubscript{1}}}
\newcommand{\myset}              {\styleLogic{s}}
\newcommand{\settwo}              {\styleLogic{s\textsubscript{2}}}
\newcommand{\setthree}              {\styleLogic{s\textsubscript{3}}}
\newcommand{\setfour}              {\styleLogic{s\textsubscript{4}}}

\newcommand{\expan}[2]          {\ensuremath{\left[#1\right]\!\styleAlgorithm{#2}}}
\newcommand{\wild}[1]              {\#\styleAlgorithm{#1}}
\newcommand{\shuf}[1]              {\ensuremath{\left(#1\right)\!\styleAlgorithm{!}}}
\newcommand{\sub}[1]              {\ensuremath{\left(#1\right)\!\styleAlgorithm{\subset}}}
\newcommand{\fuzz}[1]              {\ensuremath{\left\langle#1\right\rangle\!\styleAlgorithm{@}}}
\newcommand{\cfuzz}[1]              {\ensuremath{\left\langle#1\right\rangle\!\styleAlgorithm{\$}}}
\newcommand{\rep}[2]               {\ensuremath{\left\vert#1\right\vert\!\styleAlgorithm{#2}}}
 
 \newcommand{\slide}[2]          {\ensuremath{\left(#1\right)\!\styleAlgorithm{#2}}}

 \newcommand{\merge}[2]          {\ensuremath{\left(#1 
: #2\right)\!\styleAlgorithm{+}}}
 \newcommand{\mulmerge}[2]          {\ensuremath{\left(
\begin{array}{c}
 #1\\ 
 #2
\end{array} 
\right)\!\styleAlgorithm{+}}}
 
\newcommand{\pre}[1]              {\styleAlgorithm{P}\ensuremath{\left(#1\right)}}

\newcommand{\s}              {\ensuremath{\;\;}}

\newcommand{\etal}{et al.}
\newcommand{\kilo}{k}
\newcommand{\bit}{bit}
\newcommand{\second}{s}
\newcommand{\percent}{\%}

\newcommand{\ccount}{\mathit{count}}
\newcommand{\predicate}{\mathit{select}}

\newcommand{\linear}{\mathit{relation}}
\newcommand{\cond}{\mathit{cond}}

\definecolor{Gray}{gray}{0.9}

\usepackage{array}
\newcolumntype{L}[1]{>{\raggedright\let\newline\\\arraybackslash\hspace{0pt}}m{#1}}
\newcolumntype{C}[1]{>{\centering\let\newline\\\arraybackslash\hspace{0pt}}m{#1}}
\newcolumntype{R}[1]{>{\raggedleft\let\newline\\\arraybackslash\hspace{0pt}}m{#1}}

\begin{document}

\author{Ahmad Ibrahim}
\affiliation{%
 \institution{CISPA Helmholtz Center for Information Security}
 \city{}
 \country{}
}\email{ahmad.ibrahim@cispa.de}
\authornote{Both authors contributed equally to the paper}

\author{Hamed Nemati}
\affiliation{%
  \institution{Stanford University \\
  CISPA Helmholtz Center for Information Security}
 \city{}
 \country{}
}\email{hnnemati@stanford.edu}
\authornotemark[1]

\author{Till Schlüter}
\affiliation{%
 \institution{CISPA Helmholtz Center for Information Security}
 \city{}
 \country{}
}\email{till.schlueter@cispa.de}

\author{Nils Ole Tippenhauer}
\affiliation{%
 \institution{CISPA Helmholtz Center for Information Security}
 \city{}
 \country{}
}\email{tippenhauer@cispa.de}

\author{Christian Rossow}
\affiliation{%
 \institution{CISPA Helmholtz Center for Information Security}
 \city{}
 \country{}
}\email{rossow@cispa.de}

\title{Microarchitectural Leakage Templates and Their Application to Cache-Based Side Channels}

\begin{abstract}
The complexity of modern processor architectures has given rise to sophisticated interactions among their components.
Such interactions may result in potential attack vectors in terms of side channels, possibly
available to userland exploits \new{to leak secret data}.
Exploitation and countering of such side channels requires a detailed understanding of the target component.
However, such detailed information is commonly unpublished for many CPUs.

In this paper, we introduce the concept of Leakage Templates to abstractly describe specific \new{side channels} and identify their occurrences in binary applications. We design and implement \framework{}, a framework to derive the generic Leakage Templates from
individual code sequences that are known to cause leakage (e.g., found by prior
work). \framework uses a combination of \textit{instruction fuzzing},  \new{\textit{instructions' operand mutation}} and \textit{statistical analysis} to explore undocumented behavior of microarchitectural optimizations and \new{derive sufficient conditions on vulnerable code inputs that if hold can trigger a distinguishing behavior}. Using \framework we identified novel leakage primitives based on Leakage Templates (for ARM Cortex-A53 and -A72 cores), in particular related to \emph{previction} (a new premature cache eviction), and  prefetching behavior. \new{We show the utility of Leakage Templates by re-identifying a prefetcher-based vulnerability in OpenSSL 1.1.0g first reported by Shin et al.~\cite{10.1145/3243734.3243736}.}
\end{abstract}

\maketitle
\section{Introduction}

The past decade has witnessed a surge in side-channel attacks that exploit underspecified  or undocumented hardware features~\cite{spectre, meltdown, attackVerf, Aciicmez:2006:TCA:2092880.2092891, zhang2012cross, Neve:2006:AAC:1756516.1756531, Tromer:2010:ECA:1713125.1713127}, mostly focusing on cache-related leakage.  The hidden nature of microarchitectural features has led to the development of techniques to
test for the presence  in hardware, and semi-automatically identify new vulnerabilities~\cite{scamv, checkmate,DBLP:conf/ndss/GrasGKBR20,DBLP:conf/se/NilizadehNN20,DBLP:conf/uss/MoghimiLS020, DBLP:journals/corr/abs-2106-03470, ragab_crosstalk_2021, DBLP:journals/corr/abs-2006-14147} that allow the attacker to violate process isolation to obtain secret data, or to manipulate the victim's execution.
Existing approaches to identify \new{side channels} commonly yield architecture-specific \emph{distinguishing examples}, i.e., concrete code examples that represent side-channel leakage. Generalization from such concrete examples is a known hard problem, as it requires a detailed understanding of the processor component that introduces the channel.
Details on information flow properties of microarchitectures  are generally scarce, not publicly available, or depend on industrial secrets.
As result, determining whether a given application is vulnerable to cache-related side-channel leakage is challenging.

\new{Recent approaches already demonstrate the power of automation in the context of side channel analysis~\cite{DBLP:conf/ndss/GrasGKBR20,DBLP:conf/uss/MoghimiLS020, DBLP:journals/corr/abs-2106-03470, ragab_crosstalk_2021, DBLP:journals/corr/abs-2006-14147}.
Gras et al.~propose ABSynthe~\cite{DBLP:conf/ndss/GrasGKBR20} to automatically infer \emph{leakage maps} that show how instructions influence each other's contention behavior.
Their system can identify and optimize for microarchitecture-specific side channels exploiting hyperthreading.
In addition, Weber et al.~introduce a fuzzing framework Osiris that synthesizes instruction sequences to identify timing-based side channels~\cite{DBLP:journals/corr/abs-2106-03470}.
To this end, Osiris proposed an automated system to identify sequences that trigger and reset certain microarchitectural states.
Finally, with Transynther~\cite{DBLP:conf/uss/MoghimiLS020}, Moghimi et al.~present an approach to synthesize Meltdown-type attacks.
Transynther varies known attack patterns to create candidate attack code snippets and then evaluates whether these snippets leak data.
All of these approaches have in common that they automate the search for new (variants of) side channels, a big leap towards automation.
Having said this, they (i) are specific towards their use case (e.g., contention, Meltdown), (ii) limit their search space (e.g., instruction operands are largely ignored), and (iii) focus on attack code generation instead of finding a generic pattern for vulnerabilities that can be matched in existing code.}

\new{In this work, we introduce the concept of \emph{Leakage Templates} (or \LT{}s for short), which consist of a generalized code sequence and a set of relations on input parameters that, when satisfied, can trigger specific leakage behavior in victim code (see~\autoref{sec:ltdef} for more details)}. Given such an \LT{} for a target hardware platform, we can identify code sequences (and required input values) that expose a specific side channel behavior (see~\autoref{sec:reidentifyingPoC} for an example).
We thereby address two research questions:
(1) How can we learn generic \LT{}s for largely-undocumented leakage behavior?
(2) How can we use the \LT{}s to find side channel vulnerabilities?

To this end, we design and implement \framework{} (source is available at~\cite{plumber}) to facilitate generating such \LT{}s, leveraging \emph{instruction fuzzing}, \new{\emph{instructions' operands mutation}}, and \emph{statistical analysis}.
The design of \framework is based on exploring the architectural space through the execution of program-input pairs, and analyzing the resulting microarchitectural states (focusing on caching). We design a domain-specific language that simplifies the generation of a large number of instruction sequences (i.e., programs) and mutating their operands. Further, we use a statistical analysis approach to classify microarchitectural states and to extract relations on inputs.
To validate our approach, we studied the cases of the cache replacement policy of the ARM Cortex-A72 and two microarchitectural features of the ARM Cortex-A53 processors: \emph{previction} and \emph{prefetching}.

Previction is a recently discovered yet widely undocumented processor behavior of evicting cache lines \emph{before} the corresponding cache set is full~\cite{scamv}.
Since previction behaves differently for two addresses that only differ in their cache lines' offset, it may violate existing assumptions used to secure software~\cite{Neve:2006:AAC:1756516.1756531}. 
Prefetching, on the other hand, is a partially documented feature that allows the processor to detect regular memory access patterns and to fill cache lines with anticipated addresses by proactively continuing the pattern.
However, many details of prefetching (e.g., the number of prefetched lines) are undocumented. In addition, the cache replacement policy of ARM processors is not well-documented.

Leveraging \framework, we analyze cache replacement policy, previction and prefetching behavior of the processor's core and derive related \LT{}s.
For the replacement policy experiment, the derived \LT{} establishes eviction strategies of L1 data cache.
In the case of previction, the extracted \LT{}s reveal conditions under which bits of memory address loads are leaked. 

For prefetching, we leverage \framework to discover parameters such as the minimal number of loads to trigger prefetching, the impact of intermediate instructions, the impact of page boundaries on prefetching, and the impact of cache hits.
In other words, we show how the obtained \LT{}s expose prefetching \new{side channels} that allow to infer the control flow of a program
and to leak secret information.
Those channels are different from existing prefetching-based attacks discussed in the literature~\cite{DBLP:conf/micro/BhattacharyaRM12,10.1145/2976749.2978356,10.1145/3243734.3243736,vicarteopening}, which target the x86 architecture and either attack a software-based prefetcher~\cite{10.1145/2976749.2978356} or use a simulated CPU~\cite{DBLP:conf/micro/BhattacharyaRM12} for their analysis.
Most similar to our approach is the work of Shin et al.~\cite{10.1145/3243734.3243736} which exploits the hardware prefetcher to attack a constant-time Elliptic Curve Diffie-Hellman (ECDH) implementation from OpenSSL.
However, compared to ours, they used a different side channel, i.e. the effect of the order of accessed lookup table entries on the behavior of the prefetcher.

Although  our focus in this paper is on the ARM architecture and cache-based side channels, the \framework design is generic and can be used to detect \LT{s} for other hardware features 
and on other architectures. Also,
while the main goal of \framework is to ease generating \LT{s}, it can also be used to help reverse engineering of undocumented  microarchitectural features. For example,~\autoref{sec:discussion:bp}
 shows how \framework{} can be applied to discover the \new{structure} of the Cortex-A53's branch predictor.

\parhead{Contributions.} Our main contributions are as follows:
\begin{itemize}
  \item We introduce the concept of \emph{Leakage Templates}, which allow to  identify code sequences (and required values) that expose a specific side channel behavior in a binary application executed on a specific architecture (\autoref{sec:leakageTemplates}).
 \item We design and implement \framework, a framework that generates \LT{}s, and allows to obtain a deeper understanding of hidden behavior of microarchitectures (\autoref{sec:reverse}, \autoref{sec:impl}, and \autoref{sec:discussion}).
 \item We show \framework's efficacy by investigating the undocumented eviction policy of Cortex-A72's L1 cache and previction and prefetching behaviors of the  Cortex-A53  processors (\autoref{sec:experiments}), and identify five novel \new{side channels} (\autoref{sec:attacks}).
 \item \new{We demonstrate how a derived LT can be used to identify a side channel in a binary application (\autoref{sec:reidentifyingPoC}).}
\end{itemize}
\section{Background}
\label{sec:background}

Side channels are hidden information flow paths, which are potentially exploitable by an attacker to leak data.
The number of attacks exploiting microarchitectural features, like caches~\cite{Tsunoo03cryptanalysisof,Aciicmez:2006:TCA:2092880.2092891, zhang2012cross,Neve:2006:AAC:1756516.1756531, Tromer:2010:ECA:1713125.1713127},  continues unabated.
Therefore, the study of information flow analysis techniques to ensure the absence of side channel leakages is a topic of increasing relevance.
Central to such an analysis is a model capturing the channel.
However, the complexity of modern processors and the lack of information about their features make infeasible to explicitly model all the relevant, complex, and intertwined features like cache hierarchies and out-of-order execution. 
Abstract \textit{observational models} tackle this problem by
over-approximating attacker capabilities to observe flow of information via side channels.

\subsection{Information Flow Analysis Tools -- \tool{}}
\label{sec:scamv}

A key requirement of observational models is their \emph{soundness}, i.e. observationally equivalent states should lead to executions that
cannot be distinguished by an attacker on real hardware. 
\tool~\cite{scamv} automates validation of observational models' soundness. At high level, Scam-V generates well-formed random binary programs, which are denoted by \prog.  It then constructs pairs of initial states $\inputzero$ s.t. executions of \prog from these states are observationally indistinguishable on the model. \tool{} then tests if the two states are also indistinguishable on hardware in the presence of undocumented microarchitectural components.
Any experiment which enables the attacker to distinguish one microarchitecture-level execution from another represents a \emph{counterexample} to the soundness of the observational model.
We call a counterexample a \emph{\counterexample{}} in this paper. A \counterexample{} can be seen as an instantiation of our Leakage Templates that can be used by an attacker to infer a function of the (secret) data via relevant side channels.

\begin{figure}
\includegraphics[width=1\linewidth]{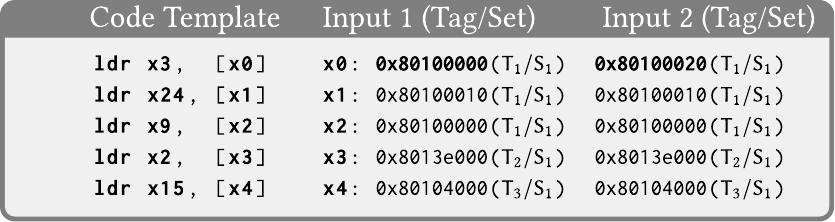}
\caption{A previction counterexample.}
\Description{A previction counterexample.}
\label{fig:counter2}
\end{figure}
\parhead{Previction.}
Previction is an undocumented behavior of the ARM Cortex-A53 processors which invalidates cache-related observational models.
Previction causes a cache line to be evicted \emph{before} the corresponding cache set is full.
Scam-V~\cite{scamv} discovers a handful instances of this behavior. However, the real cause of \emph{previction} is unknown. 
The authors conjecture that the processor detects a short sequence of loads to the same cache set and anticipates more loads to the same set with no reuse of previously loaded values. It evicts the valid cache line in order to make space for more colliding lines.

\autoref{fig:counter2} depicts a previction counterexample.
The program consists of five loads. The given inputs are observationally equivalent: they only
differ for the value of \lstinline|x0|, which affects the address used for the first load. However, the addresses \lstinline|0x80100000| and \lstinline|0x80100020| in  \lstinline|x0| have the same tag and cache
set index and only differ in the offset within the same cache line.
The addresses of all load instructions are mapped to the same cache set, i.e., set $0$. Since the cache is $4$-way associative and is initially empty, one expects no eviction to occur. Executing the program with the given inputs on the real hardware, however, results in two different cache states. In one case \lstinline|x0| is present in the cache, while in the other case it is not.

\subsection{ARMv8 internal memory subsystem}\label{sec:background:rpi}
ARMv8 processors have two levels of caches:
(1)~a level one (L1) cache per each core; and
(2)~a last level cache (L2) shared between cores.
When the CPU needs to read a memory location that is currently not cached, it fetches the requested data from memory into a cache line and tags that line with the memory location where the data was read from.
When a line is loaded from memory and all potential destination lines in the cache are occupied, the CPU uses a specific replacement policy to decide which colliding line(s) should be evicted. The eviction policy ensures that the cache always stores the most popular content, thus using the available space efficiently.

\autoref{tab:notaitons} shows our notation to extract cache-related information from an address. \texttt{sameTag}, \texttt{sameSet} \new{and \texttt{samePage}} are predicates checking for equality of cache tag, set, \new{or page indices} of addresses.
%

\parhead{Prefetcher.}
The L1 cache implements a prefetcher, for some configurable $k \in \mathbb{N}$. When the prefetcher detects $k$ cache misses whose set indices are separated by a fixed stride, the prefetcher starts to fill the cache with a sequence of lines from memory locations whose addresses match the stride of the initial cache misses.
We call such sequences \emph{prefetch streams}.
An exception happens when prefetching crosses a small page (4K) boundary. In this case the prefetcher stops fetching data from the adjacent page.

\begin{table}
\centering
\caption{Summary of notations.}
\label{tab:notaitons}
\resizebox{0.7\columnwidth}{!}{
\begin{tabular}{ll}
\hline
\textbf{Notation}    &  \textbf{Description} \\
\hline
\addr{i}                                     & A physical address                   \\
\bits{m}{n}{\addr{i}}                        & Bits $m$ through $n$ of  \addr{i}    \\
$\exidx{\addr{i}} = \bits{6}{12}{\addr{i}}$  & \addr{i} cache set index           \\
$\extag{\addr{i}} = \bits{13}{31}{\addr{i}}$ & \addr{i} cache tag                 \\
$\exword{\addr{i}} = \bits{2}{5}{\addr{i}}$  & \addr{i} word offset               \\
$\exbus{\addr{i}} = \bits{4}{5}{\addr{i}}$   & \addr{i} bus round                 \\
\new{$\expage{\addr{i}} = \bits{12}{31}{\addr{i}}$} & \new{\addr{i} page index}
\\
$\tageq{{\addr{i},\addr{j}, \dots}}$         & Cache tag equality predicate       \\
$\seteq{{\addr{i},\addr{j}, \dots}}$         & Cache set equality predicate       \\
$\pageeq{{\addr{i},\addr{j}, \dots}}$         & Page index equality predicate       \\
\hline
\end{tabular}}
\end{table}

\section{Leakage Templates}
\label{sec:leakageTemplates}

We now introduce \emph{Leakage Templates} and motivate their utility. 

\subsection{Goal and Motivation for Leakage Templates}

\parhead{Goal of the analyst.} The overall setting is depicted in \autoref{fig:framework}.
We assume that the analyst has concrete examples of (artificial) code (for a specific hardware architecture) that behave distinctively \new{on the microarchitectural level} for different inputs, thus exposing an undocumented behavior. We call these examples \emph{distinguishing examples}. \new{\framework{} takes as input an abstract description of a leaking code snippet in terms of a \emph{\request{}} (\requestShort) (see \autoref{sec:gts} for more details)}. The  goal of the analyst is to utilize this behavior to leak information from a real-world application. To achieve this, the analyst needs to identify code segments in the target application that \emph{under certain analyst-controllable conditions} trigger the undocumented microarchitectural behavior.

\parhead{Motivation for Leakage Templates.} As the analyst starts with concrete code sequences (in form of distinguishing examples), it is unlikely that the exact same code sequences will appear in another target application. Therefore, we need to abstract from the concrete code sequences and find: (1)~a generalized code sequence, (2)~a set of relevant attributes, and (3)~relations between those attributes that expose the specific \new{side channel}. We call such information a \emph{Leakage Template} or \emph{\LT{}}. The \LT{} abstractly defines the conditions for code segments under which the leakage is observed.
Given an \LT{}, an analyst can identify code segments in a target application that expose a side channel. 
\new{We demonstrate in \autoref{sec:reidentifyingPoC} how binaries can be scanned for code sections that match the code pattern from a \LT{}. In addition, we show how dynamic binary analysis techniques can be used to analyze a matching code section for the presence of side-channel leakage based on the relations from a \LT{}.}

\parhead{Sources of distinguishing examples.} There are at least two general ways of finding distinguishing examples. Such examples can be derived from abstract (natural language) descriptions of the architecture's behavior, e.g., in manuals. Moreover, one can extract these examples from concrete code traces that expose the intended \new{side channel}, e.g., out of tools like \tool{} or zero-day exploits.
\subsection{Motivating Example I: Caching}
Assume that simple caching behavior of a CPU was insufficiently documented, and the analyst tries to understand in which parts of a target application (and under which conditions) caching occurs.

\parhead{Initial Information.} A starting \emph{\counterexample{}} (i.e.
a distinguishing code example with a pair of inputs) could contain a sequence of two load instructions (for variable addresses), and two instances (i.e., pairs of input addresses) where different behavior was observed. In the first instance, both loads refer to the same address, while in the second instance, two distinct addresses (on different cache lines) are accessed. When executing the two instances on a clean cache state, the second load will be considerably faster for the first instance compared to the second one.

\parhead{Leakage Template.} An abstract \LT{} that describes this \new{side channel} would specify:
(1)~an abstract code template (i.e., at least two loads from symbolic addresses with potentially other instructions in between);
(2)~the two possible behaviors (i.e., fast and slow execution time of the second load); and
(3)~the relations over the loaded addresses which lead to the respective behaviors.

\parhead{Benefit of the Leakage Template.}
The starting \counterexample{} just provides one concrete code instance that leads to the leakage behavior, while the \LT{} ideally covers every possible sequence of instructions in which two loads lead to caching. Consequently, one would expect that the \counterexample{} alone will not enable the analyst to identify instances of the side channel leakage in an arbitrary target program, while the \LT{} is expected to have much higher chances to discovering related code fragments. 

\subsection{Motivating Example II: Previction}
For previction, the leakage behavior appears to require a more complex code sequence (see \autoref{sec:scamv}).

\parhead{Initial information.} In this case, a starting \counterexample{} could contain a sequence of five loads, and two sets of addresses which lead to different behaviors, as summarized in \autoref{fig:counter2}.

\parhead{Leakage Template.} An \LT{} that describes this \new{side channel} would specify: 
(1)~an abstract code template (e.g., at least five loads from symbolic addresses with potentially a number of other instructions in between);
(2)~the two possible behaviors (i.e., previction or no previction); and
(3)~the abstract relations over the loaded addresses which lead to the respective behaviors.

\subsection{Definition of Leakage Templates}
\label{sec:ltdef}
\begin{figure}
\includegraphics[width=1\linewidth]{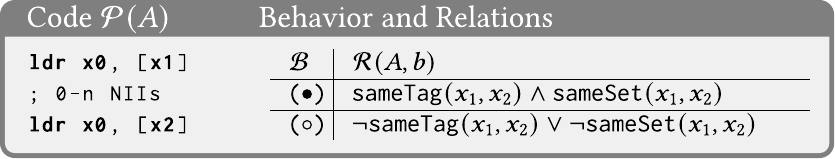}
\caption{Caching \LT{}. $\bullet$ (fast), $\circ$ (slow) are distinguishing behaviors. NIIs are Non-Interfering Instructions.}
\Description{A Leakage Template describing Caching, consisting of a code section and a section that maps the two different behaviors (fast/slow or cached/uncached) to relations.}
\label{fig:ltexample}
\end{figure}

Based on the provided motivation for \LT{}s, we now describe the components of \LT{}s themselves. An \LT{} is a triple $(\proghat, \behavior, \relation)$; with $\proghat$ being a \emph{code template} with a set of attributes ${A}\subseteq \mathbb{A}$, $\behavior$ a set of observed distinct behaviors, and $\styleSet{R}: \mathbb{A} \times \behavior \to 2^{\mathbb{P}_{{A}}}$ maps a behavior $b \in \behavior$ to a (set of) predicate(s) on attributes, where $\mathbb{P}_{{A}}$ is the set of predicates on ${A}$.
We note that our definition of behavior is generic. For instance, it may refer to temporal (e.g., measuring execution time) or spatial (e.g., monitoring cache content) behavior. For the latter, the behavior refers to the difference between the initial (before execution of the code template) and final (after execution of the code template) state of the monitored component (e.g., cache).
Given this description, we provide additional details on the resulting \LT{} for our caching example in \autoref{fig:ltexample}. In~\autoref{sec:previction} we show how an \LT{} for the previction example can be derived.

\section{\framework}
\label{sec:reverse}
We now present our design of \framework, a framework to automatically derive \LT{}s (focusing on caching behavior). In particular, in our design we had to address the following challenges:

\begin{itemize}
  \item \textbf{C1}: We need to construct efficient specifications to steer testcase generation towards generating inputs which are  likely to trigger a specific microarchitectural behavior. 
  \item \textbf{C2}: Size of input space (number of possible code sequences and input values) is too large. Thus, we need an approach to explore the input space and  ensure high test coverage which in turn increases accuracy of the derived relations. 
  \item \textbf{C3}: We need to get accurate measurements with minimal noise, close to `ground truth' of the respective channel.
  \item \textbf{C4}: The relations between attributes in code sequences and triggered behaviors can be very complex and counter-intuitive, and manual derivation of such relations (if not impossible) is highly error prone. So, we need to develop a statistical analysis technique to automate finding the correlation between attributes and observed behavior.
  \end{itemize}

\subsection{Abstract Framework Design}
\label{sec:gts}

\framework{'s} input is a \emph{\request} (\requestShort), which  abstractly describes programs to analyze in a domain-specific language, and whose leakage effects are to be monitored. The detailed description of \requestShort\xspace and the language  (and how it addresses \textbf{C1}) is provided in  \autoref{sec:requests} and \ref{sec:language}.
The framework  outputs the \LT{}, including the behavior of the monitored microarchitectural components, and the relations between attributes and the behaviors.

\begin{figure}
	\centering
	\includegraphics[width=0.95\linewidth]{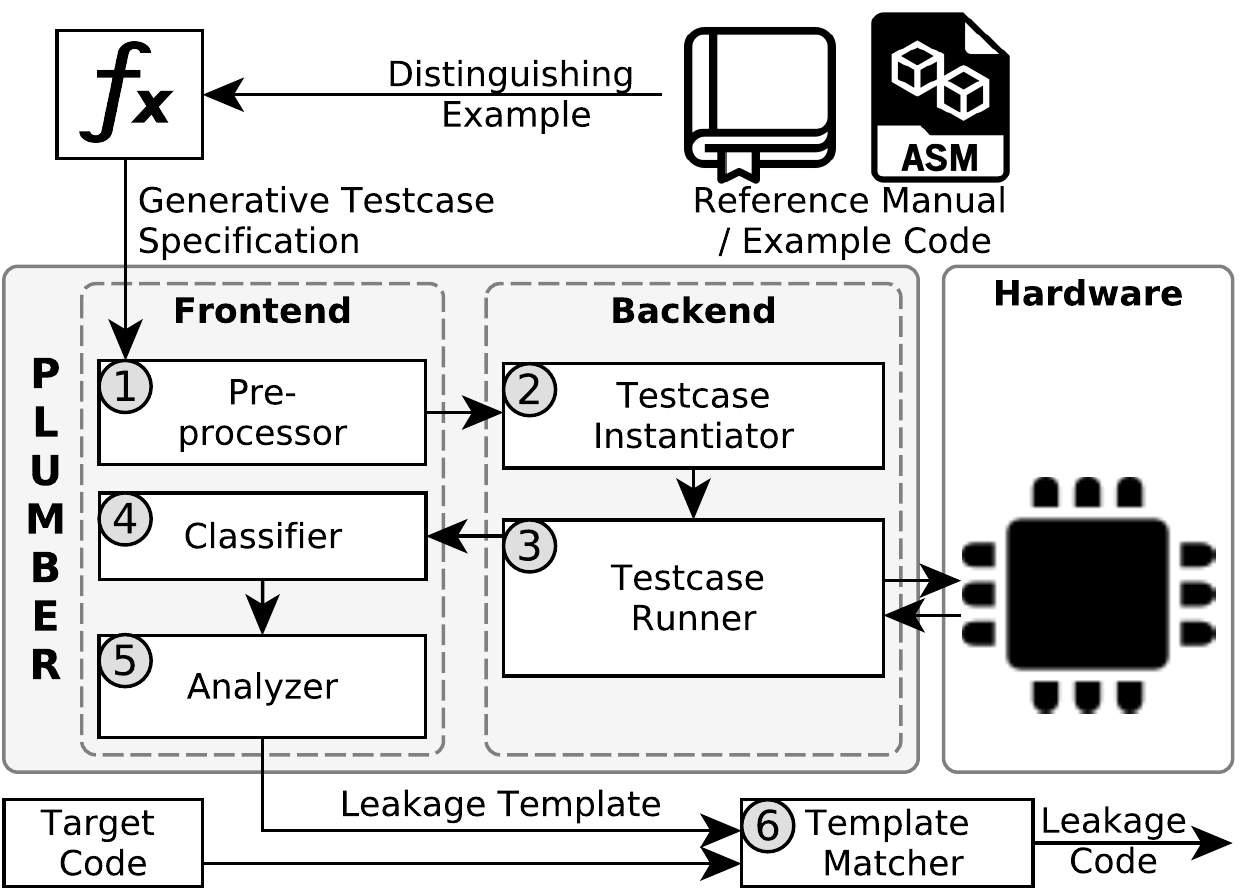}
	\caption{\new{Overview of \framework{} components.}}
  \Description{Overview figure presenting the \framework{} framework. Input to the framework is a \request, which is produced based on an distinguishing example (e.g., derived from a reference manual or from example code (leakage witness)). The framework itself is split into two components: Frontend and Backend. The \requestShort{} first passes through the Preprocessor (1, Frontend), is then passed to the Testcase Instantiator (2, Backend) and then to the Testcase Runner (3, Backend). The Runner interacts with the hardware. After execution, the resulting data is passed on to the Classifier (4, Frontend), then to the Analyzer (5, Frontend). The output of the \framework{} framework is a Leakage Template, which can be fed into a Template matcher (6), together with a target code sample. The goal of the template matcher is to identify leaking code sequences in the target code.}
	\label{fig:framework} 
\end{figure}

As shown in \autoref{fig:framework}, the framework consists of two parts.
The \emph{Backend} instantiates and executes testcases (i.e., program-input pairs)
from a (preprocessed) \requestShort. 
The \emph{Frontend} encapsulates the Backend, towards the outside it receives a \requestShort\xspace and outputs \LT{}s.

We present each component by describing the steps required to derive an \LT{} from a \requestShort.
\circles{1} The \emph{Preprocessor} parses the \requestShort\xspace
and forwards the result to the \emph{Testcase Instantiator}. \circles{2} The Instantiator then systematically generates testcases (addressing \textbf{C2})
based on the parsed preprocessed \requestShort\xspace.
\circles{3} The \emph{Runner} executes every testcase in a controlled environment (addressing \textbf{C3})  and returns the behavior 
to the \emph{Classifier}. \circles{4} The Classifier classifies the programs (if necessary) based on their behavior and stores them. \circles{5} The \emph{Analyzer} interprets classified behaviors (addressing \textbf{C4}) and returns relations on the attributes which trigger specific microarchitectural behavior (i.e., $\behavior$ and $\relation$ of the \LT{}).

A possible application of an \LT{} is to identify instances of it in binaries. This can be done using a \emph{Template Matcher} \circles{6}, which may use static and dynamic binary analysis techniques to find code sections matching the \LT{'s} code pattern as well as its relations. We discuss a proof-of-concept implementation in \autoref{sec:reidentifyingPoC}.

\subsection{Definition of \requestShort}
\label{sec:requests}
The \requestShort\xspace (used as input of \framework) intuitively defines sequence(s) of instructions to be executed over (\new{mutated}) operands, and microarchitectural component(s) to be monitored (e.g., content of the cache).
The \requestShort\xspace can also specify the initial state of (monitored)  component(s) before the execution of its instruction sequences.

To address \textbf{C1} as outlined earlier, we  introduce a domain-specific generative language. The language provides three main features. (i) It allows us to abstractly specify possible mutations of the program (i.e. enabling very generic code templates), (ii) it allows us to specify which data in the programs should be fuzzed \new{or mutated} (i.e. defining domains for attributes), and (iii) it allows us to specify a set of relations to apply to non-fuzzed data in the testcases.

\subsection{Specification Language}
\label{sec:language}
A \request\ (\requestShort)\ is formed of a sequence of directives 
that specify different operations, e.g., arithmetic operations or \NOP instructions.
The proposed language defines an extensible set of directives, which includes
$\mem$, $\arith$ and $\nop$.
The directive \mem denotes memory loads, the directive \arith denotes arithmetic or logical operations, and the directive \nop denotes \NOP instructions.
Additionally, the language defines two directives that allow reverse engineering the branch predictor (see \autoref{sec:discussion}).
\setbranchtwo{\bidone}{\bool} denotes an instruction that sets a variable (identified by \bidone) to a boolean value \bool, and 
\branchthree{\bidone}{\bool}{\bStep} denotes a branch instruction that jumps \bStep{} steps when the value of the variable identified by \bidone is \bool. The default value of a variable is false (\bfalse).

The addresses of load operations are mapped to a certain cache set and have specific tags.
It is possible to define the tag and set attributes of a memory directive.
We use \memtwo{\tagone}{\setone} to refer to a load from an address with tag \tagone\ and cache set \setone. The values of these attributes allow determining the relation between tags or sets of different loads.
When omitted, these attributes acquire default values which are identical for the same request.
The language allows defining arithmetic relations between the tag (and set) attributes of different memory directives.
For example, the \requestShort\ $\mem\;\mem\;\memtwo{\tagone}{\setone}\;\memtwo{\tagone}{\setone}\;\memtwo{\tagone+1}{\setone+5}\;\arith\;\arith\;\nop$ represents two loads of memory addresses with the same set and the same tag (default), followed by two other loads with a different tag \tagone{} and a different set \setone{} and a fifth load with tag ($\tagtwo= \tagone+1$) and set ($\settwo=\setone+5$), two arithmetic or logical operations, and a \NOP instruction.
The values of the operands of other instructions, e.g., arithmetic operations, can be defined in the same manner, e.g., \arithtwo{\valueone}{\valuetwo} represents an arithmetic operation with two operands \valueone and \valuetwo.

The language also provides an extensible set of macros and
operators that allow constructing a meaningful \requestShort{} as well as
defining the initial state of hardware components. Note that this language is extensible,
i.e., it allows defining new macros and operators for investigating
various processor components.

\parhead{\bf Power \normalfont \expan{}{\dirone{\attr}, n, i}\textbf{.}} This macro allows repeating directive(s) $n$ times, while incrementing the attribute \attr\ of the directive \dir\ by a value of $i$.
For example, the \requestShort\ $\expan{\memtwo{\tagone}{\setone}}{\memone{\myset}, 2, 1}$ can be used to refer to the \requestShort: $\memtwo{\tagone}{\setone} \s\memtwo{\tagone}{\setone+1}$.
The power macro can be also used with a single input $n$. In this case the directive(s) are repeated $n$ times. For example, the \requestShort\ presented above can also be expressed as: $\expan{\mem}{2}\s \expan{\memtwo{\tagone}{\setone}}{2} \s\expan{\arith}{2}\s \nop$.

\parhead{\bf  Wildcard \wild{n}.} This macro expands to $n$ arbitrary directives that do not perform memory operations. For example, one possible expansion of $\mem\s \wild{3}\s\mem$ is $\mem\s\nop\s\nop\s\nop\s\mem$.

\parhead{\bf  Shuffle \shuf{}.}
This operator generates all possible permutations of a \requestShort\ while omitting those with similar directives. For example,
\shuf{\expan{\mem}{2}\s \memtwo{\tagone}{\setone}} refers to the set: $\{\mem\;\; \mem\;\; \memtwo{\tagone}{\setone}; \mem\s\memtwo{\tagone}{\setone}\; \mem;\ \memtwo{\tagone}{\setone}\; \mem\s \mem\}$.

\parhead{\bf  Subset \sub{}.}
This operator generates all possible subsets of a \requestShort\ while omitting those with similar directives. For example,
\sub{\expan{\mem}{2}\s \memtwo{\tagone}{\setone}} refers to the set: $\{\mem\s \mem;$ $ \mem\s\memtwo{\tagone}{\setone};\ \mem;\ \memtwo{\tagone}{\setone}\}$.

\parhead{\bf  Slide \slide{}{n}.}
For a given \requestShort, this operator shifts all loaded addresses one set at a time up to $n$ times. 
For example, \slide{\memtwo{\tagone}{\setone}\s\memtwo{\tagtwo}{\settwo}}{3} refers to the set: $\{\memtwo{\tagone}{\setone}\s\memtwo{\tagone}{\setone};\ \memtwo{\tagone}{\setone+1}\s\memtwo{\tagone}{\setone+1};\ \memtwo{\tagone}{\setone+2}\s\memtwo{\tagone}{\setone+2}\}$.

\parhead{\bf  Merge \merge{}{}.} This operator merges two requests by sliding the directives of the first over the second as demonstrated by the following example. \merge{\memtwo{\tagone}{\setone}\s\memtwo{\tagtwo}{\settwo}}{\memtwo{\tagthree}{\setthree}\s\memtwo{\tagfour}{\setfour}} refers to the set:

$\left\lbrace\begin{aligned}
\memtwo{\tagone}{\setone}\;\memtwo{\tagtwo}{\settwo}\s\memtwo{\tagthree}{\setthree}\s\memtwo{\tagfour}{\setfour};\ \memtwo{\tagone}{\setone}\s\memtwo{\tagthree}{\setthree}\s\memtwo{\tagtwo}{\settwo}\;\memtwo{\tagfour}{\setfour};\\
\memtwo{\tagthree}{\setthree}\;\memtwo{\tagone}{\setone}\s\memtwo{\tagfour}{\setfour}\s\memtwo{\tagtwo}{\settwo};\ \memtwo{\tagthree}{\setthree}\s\memtwo{\tagfour}{\setfour}\s\memtwo{\tagone}{\setone}\;\memtwo{\tagtwo}{\settwo}
\end{aligned}\right\rbrace$

\parhead{\bf Load offset \new{mutation} \fuzz{}.}
For every load instruction, the operator signals generation of a testcase for all possible addresses with the indicated tag and set, i.e., it brute forces word offsets. For example, \fuzz{\mem\s\mem} generates a set formed of all two loads having the same tag and set with all possible combinations of word offsets.

\parhead{\bf Cache line \new{mutation} \cfuzz{}.}
For every load instruction, this operator signals the generation of a testcase for every possible memory address having the indicated tag and word offset, i.e., it brute forces all possible sets. For example, \cfuzz{\mem\s\mem} generates a set formed of all two loads that have the same tag for all possible combinations of sets, i.e., \{\memtwo{\tagone}{\setone}\s\memtwo{\tagone}{\settwo}\} for every set \setone and \settwo.

\parhead{\bf  Repetition \rep{}{n}.}
This operator repeats the \requestShort\ $n$ times, e.g., the \requestShort\ \rep{\mem\s\memtwo{\tagone}{\setone}}{3} corresponds to: $\{\mem\s\memtwo{\tagone}{\setone}\ ;\;\mem\s\memtwo{\tagone}{\setone}\ ;\;\mem\s\memtwo{\tagone}{\setone}\}$.

\parhead{\bf  Precondition \pre{}.}
This operator allows setting up the state of different hardware components before the execution of testcase. For instance, the \requestShort\ \pre{\memtwo{\tagone}{\setone}\s\memtwo{\tagtwo}{\setone}} \cfuzz{\mem\s\mem} generates cache line \new{mutation} testcases where two lines in \setone\ are already occupied.

\section{Design \& Implementation}
\label{sec:impl}

\framework{} currently targets the ARM architecture. It is implemented in \texttt{C} and \texttt{Python} as well as \texttt{ARM assembly}. We exploit \texttt{ARM assembly} for: (a)~implementing testcases, (b)~setting up and (accurately) reading architectural components, and (c)~increasing the performance. However, our design is applicable to other architectures. 
In the following, we present the implementation details for
\framework's components presented in \autoref{fig:framework}.

\subsection{Frontend}
\label{sec:impl:frontend}
The Frontend handles user interaction and has three  components:

\parhead{Preprocessor.} 
The preprocessor interprets a given \requestShort{} and forwards the results to the Backend. It expands repetition, power and wildcard macros. Wildcard is expanded by randomly picking directives from the set of defined operations. The Preprocessor is also responsible for expanding the shuffle, subset, slide and merge operators, i.e., it generates permutations of a given \requestShort{} by applying the respective operations and forwards them to the Backend.

\newcolumntype{g}{>{\columncolor{Gray}}c}
\begin{figure}[t] 
\centering
\scalebox{0.8}{
\footnotesize{
\begin{tabular}{l c c c c c} 
\toprule 
& \multicolumn{5}{c}{\textbf{Load Register}} \\ 
\cmidrule(l){2-6} 
\textbf{Output} & \cellcolor{gray} \lstinline|x0| & \cellcolor{gray}\lstinline|x1| & \lstinline|x2| & \lstinline|x3| & \lstinline|x4|\\ 
\midrule 

$1$ & \footnotesize{$1 \cdots 000000000{\mathrm{\mathbf{\underline{0}\underline{0}00}}}{\color{gray}00}$} & \footnotesize{$1 \cdots 000000000{\mathrm{\mathbf{\underline{0}\underline{0}00}}}{\color{gray}00}$} & $\cdots$ & $\cdots$ & \footnotesize{$1 \cdots 100000000{\mathrm{\mathbf{{0}{0}00}}}{\color{gray}00}$} \\
$2$ & \footnotesize{$1 \cdots 000000000{\mathrm{\mathbf{\underline{0}\underline{0}01}}}{\color{gray}00}$} & \footnotesize{$1 \cdots 000000000{\mathrm{\mathbf{\underline{0}\underline{0}00}}}{\color{gray}00}$} & $\cdots$ & $\cdots$ & \footnotesize{$1 \cdots 100000000{\mathrm{\mathbf{{0}{0}00}}}{\color{gray}00}$} \\
$3$ & \footnotesize{$1 \cdots 000000000{\mathrm{\mathbf{\underline{0}\underline{0}10}}}{\color{gray}00}$} & \footnotesize{$1 \cdots 000000000{\mathrm{\mathbf{\underline{0}\underline{0}00}}}{\color{gray}00}$} & $\cdots$ & $\cdots$ & \footnotesize{$1 \cdots 100000000{\mathrm{\mathbf{{0}{0}00}}}{\color{gray}00}$} \\
$4$ & \footnotesize{$1 \cdots 000000000{\mathrm{\mathbf{\underline{0}\underline{0}11}}}{\color{gray}00}$} & \footnotesize{$1 \cdots 000000000{\mathrm{\mathbf{\underline{0}\underline{0}00}}}{\color{gray}00}$} & $\cdots$ & $\cdots$ & \footnotesize{$1 \cdots 100000000{\mathrm{\mathbf{{0}{0}00}}}{\color{gray}00}$} \\
\rowcolor{Gray}
$5$ & \footnotesize{$1 \cdots 000000000{\mathrm{\mathbf{\underline{0}\underline{1}00}}}{\color{gray}00}$} & \footnotesize{$1 \cdots 000000000{\mathrm{\mathbf{\underline{0}\underline{0}00}}}{\color{gray}00}$} & $\cdots$ & $\cdots$ & \footnotesize{$1 \cdots 100000000{\mathrm{\mathbf{{0}{0}00}}}{\color{gray}00}$} \\
\rowcolor{Gray}
$6$ & \footnotesize{$1 \cdots 000000000{\mathrm{\mathbf{\underline{0}\underline{1}01}}}{\color{gray}00}$} & \footnotesize{$1 \cdots 000000000{\mathrm{\mathbf{\underline{0}\underline{0}00}}}{\color{gray}00}$} & $\cdots$ & $\cdots$ & \footnotesize{$1 \cdots 100000000{\mathrm{\mathbf{{0}{0}00}}}{\color{gray}00}$} \\
\rowcolor{Gray}
$7$ & \footnotesize{$1 \cdots 000000000{\mathrm{\mathbf{\underline{0}\underline{1}10}}}{\color{gray}00}$} & \footnotesize{$1 \cdots 000000000{\mathrm{\mathbf{\underline{0}\underline{0}00}}}{\color{gray}00}$} & $\cdots$ & $\cdots$ & \footnotesize{$1 \cdots 100000000{\mathrm{\mathbf{{0}{0}00}}}{\color{gray}00}$} \\
\rowcolor{Gray}
$8$ & \footnotesize{$1 \cdots 000000000{\mathrm{\mathbf{\underline{0}\underline{1}11}}}{\color{gray}00}$} & \footnotesize{$1 \cdots 000000000{\mathrm{\mathbf{\underline{0}\underline{0}00}}}{\color{gray}00}$} & $\cdots$ & $\cdots$ & \footnotesize{$1 \cdots 100000000{\mathrm{\mathbf{{0}{0}00}}}{\color{gray}00}$} \\
$9$ & \footnotesize{$1 \cdots 000000000{\mathrm{\mathbf{\underline{1}\underline{0}00}}}{\color{gray}00}$} & \footnotesize{$1 \cdots 000000000{\mathrm{\mathbf{\underline{0}\underline{0}00}}}{\color{gray}00}$} & $\cdots$ & $\cdots$ & \footnotesize{$1 \cdots 100000000{\mathrm{\mathbf{{0}{0}00}}}{\color{gray}00}$} \\
&&&$\cdots$ &&\\
$2^{20}$ & \footnotesize{$1 \cdots 000000000{\mathrm{\mathbf{\underline{1}\underline{1}11}}}{\color{gray}00}$} & \footnotesize{$1 \cdots 000000000{\mathrm{\mathbf{\underline{1}\underline{1}11}}}{\color{gray}00}$} &$\cdots$ & $\cdots$ & \footnotesize{$1 \cdots 100000000{\mathrm{\mathbf{{1}{1}11}}}{\color{gray}00}$} \\
\bottomrule 
\end{tabular}
}}
\caption{\new{Input bit table from a class causing previction. Shaded columns are registers responsible for previction. Shaded rows are inputs missing from the table. Mutated bits are in bold and the bits causing previction are underlined.}}
\Description{A bit table.}
\label{tbl:bittbl} 
\end{figure}

\parhead{Classifier.} 
This component classifies the output of the Backend based on the  behavior of the monitored component, e.g., for previction/prefetching, output is classified based on the occurrence of previction/prefetching (or the previcted/prefetched addresses).
For every behavior, the classifier generates a \emph{bit table} containing the  \emph{binary representation} of \new{mutated} instruction operands, e.g., accessed addresses for cache line \new{mutation}. 
\new{\autoref{tbl:bittbl} shows an example of a bit table generated for previction.}
Each testcase is represented by one row in the table, each column represents a  loaded address.

\begin{figure}
\centering
\includegraphics[width=1\linewidth]{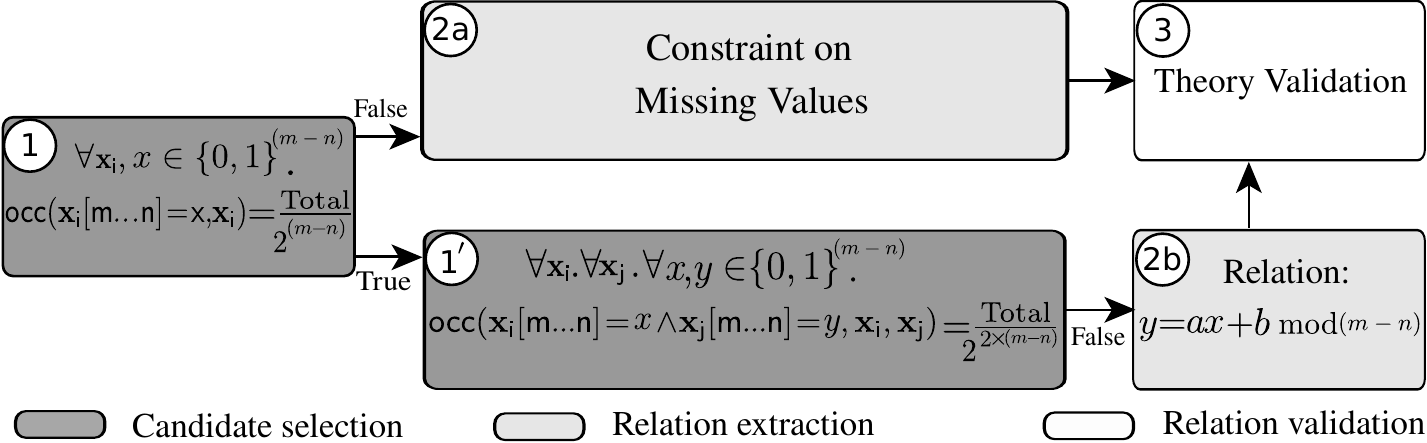}
\caption{The analyzer function flowchart for previction.}
\Description{A flowchart visualizing how the analyzer candidate selection and relation extraction works. The formulas used in this figure and the flow described here correspond to the textual description of the Analyzer in section \ref{sec:impl:frontend}.}
\label{fig:analyzer}
\end{figure}

\parhead{Analyzer.} 
The analyzer extracts  relations between inputs and their effect on the monitored microarchitectural behavior of the component. 
It exposes a set of primitive operations over bit tables to generate and validate such relations. Our primitives have similar meaning as SQL statements. Examples include:
(a)~$\ccount()$ which counts the number of rows (or columns) in a table;
(b)~the variadic function $\predicate(\mathit{cond}(\textbf{\lstinline|x|}_i[m_i\dots n_i], ...), \textbf{\lstinline|x|}_i[m_i\dots n_i], ...)$ that returns all rows  whose fields are in the relation  $\cond{}$, e.g., $\predicate(\textbf{\lstinline|x|}_i[j] = 1, \textbf{\lstinline|x|}_i)$ returns rows where the $j$th bit of the register $\textbf{\lstinline|x|}_i$ is 1;
(c)~the variadic function $\linear(\textbf{\lstinline|x|}_i[m_i\dots n_i], \textbf{\lstinline|x|}_j[m_j\dots n_j], \dots)$ which takes two or more inputs and returns a linear relation over specific bits of the inputs.
We define an example analyzer function  for previction. Other analyzer functions typically follow the same strategy.
As shown in \autoref{fig:analyzer}, this function has three phases:

\emph{Candidate selection.}
This step pinpoints parts of the inputs (i.e., specific bits from every address) that are correlated with the observed behavior (e.g., previction).
Let $\circ$ denote function composition; $\nocc = \ccount\circ\predicate$ defines the composition of $\ccount$ and $\predicate$. The analyzer determines the candidate bits as follows:

For every address $\textbf{\lstinline|x|}_i$, the analyzer uses $\nocc(\textbf{\lstinline|x|}_i[m\dots n]= x, \textbf{\lstinline|x|}_i)$ to find the number of occurrences of each possible value $x \in \{0,1\}^{(m-n)}$ for the \emph{non-constant bit sequence} indexed by $m$ and $n$, i.e., mutated bits (step \circles{1}).
It compares these actual occurrences to the expected number of occurrences of this value, i.e., it checks whether $\nocc(\textbf{\lstinline|x|}_i[m\dots n]= x, \textbf{\lstinline|x|}_i) = \frac{\total}{\npos}$; where $\total$ denotes the size of the bit table; and $\npos = 2 ^ {(m-n)}$ is the number of different possible values for the bit sequence indexed by $m$ and $n$. Every address where the number of occurrences of (some or all) values deviates from the expectation is marked as a \emph{candidate address}.

If all addresses have the same number of occurrences equal to $ \frac{\total}{\npos}$, the analyzer proceeds to step \circles{\raisebox{-.9pt}{\resizebox{0.6em}{!}{1'}}}. 
In this step, the analyzer repeats the previous check on pairs of operands. 
For every pair of operands $\textbf{\lstinline|x|}_i$ and $\textbf{\lstinline|x|}_j$ the analyzer checks whether 
$\nocc(\textbf{\lstinline|x|}_i[m\dots n]= x \wedge \textbf{\lstinline|x|}_j[m\dots n]= y, \textbf{\lstinline|x|}_i, \textbf{\lstinline|x|}_j) = \frac{\total}{\nppos}$, for all possible values $x, y \in \{0,1\}^{(m-n)}$, where $\nppos = 2 ^ {2\times(m-n)}$ is the number of possible values for the two bit sequences.
Every pair of addresses for which the number of occurrences of (some or all) values deviates from the expectation is marked as a pair of \emph{candidate interrelated addresses}.

\emph{Relation extraction.}
\new{Next, the analyzer detects the constraints on certain bits in a candidate address or the interrelation between bits in candidate interrelated addresses. 
In step \circles{\resizebox{0.6em}{!}{2a}}, the analyzer uses $\nocc$ to determine these constraints by checking every candidate address for the missing values. 
In step \circles{\resizebox{0.6em}{!}{2b}}, the analyzer uses the function $\linear(\textbf{\lstinline|x|}_i[m\dots n], \textbf{\lstinline|x|}_j[m\dots n])$ to find the interrelation between the non-constant bits of interrelated candidate addresses. It finds $a$ and $b$ in the equation $y=ax+b\!\mod(m-n)$, where $x$ and $y$ represent the interrelated bits in the two addresses, respectively.}

\emph{Relation validation.}
\circles{3} This phase validates the generated constraints and relations. 
A constraint is validated by using the $\nocc$ function to check whether all value combinations of unrelated bits occur in the bit table.
A relation is validated by using $\nocc("y=ax+b", \textbf{\lstinline|x|}_i[m\dots n], \textbf{\lstinline|x|}_j[m\dots n])$ to check whether: (i)~every row in the bit table satisfies the extracted relation and (ii)~every value combination of unrelated bits occurs in the bit table.

\subsection{Backend}
The Backend is responsible for generating concrete testcases (a set of programs and their inputs) from a \requestShort{}, setting up microarchitectural components, executing the programs over their respective inputs on real hardware, and returning the microarchitectural behavior of this execution. The Backend is formed of two components:

\parhead{Instantiator.}
This component receives an expanded \requestShort{} from the Preprocessor and generates concrete testcases.
It uses an instruction store to pick instructions for each operation when generating programs. It further uses an address store to generate inputs. 
In particular, for the first occurrence of every tag and/or set attribute value, the generator picks random values from the store and queries their corresponding address. For every consecutive occurrence of a tag or set, the generator uses the previously selected tag and/or set from the first occurrence.
In the case of arithmetic relation between these attributes, the generator searches for addresses that satisfy this relation. Alternatively, the Testcase Instantiator can generate new addresses and add them to the address store.
Finally, the Instantiator generates and/or \new{mutates} inputs of instructions
as requested. 
For example, when the \requestShort{} includes an offset \new{mutation} operator, the generator generates inputs with all possible address combinations by brute forcing their word offset. The generated testcases
are forwarded to the Runner.
When the cache line \new{mutation} operator is used, the Instantiator generates inputs with all possible address combinations by brute forcing the set index.

\parhead{Runner.}
The Testcase Runner receives a testcase
from the Instantiator to be executed on the hardware. The Runner first inserts memory barriers between the program's instructions. It then connects to the hardware,
and refreshes the microarchitectural state, e.g., clears the cache.
The Runner then sets up microarchitectural components by executing the precondition part of the program. 
The program is then executed using ARM TrustZone.
ARM TrustZone provides the highest level of privilege that allows the execution of all possible instructions as well as the inspection of microarchitectural states. Most importantly, TrustZone provides direct access to the cache state through privileged debug instructions.
Having said that, \framework may also leverage other techniques to infer the microarchitectural state (e.g., Reload+Time typically used in cache side-channel attacks~\cite{flushreload, primeprobe} to infer the content of the cache) if such special debug instructions are  unavailable.

We conduct our experiments as bare-metal code---there are no background processes or interrupts which could induce noise in terms of cache content or timing. We still experience a low amount of noise due to the shared memory subsystem (such as the GPU) and because our experiments are not synchronized with the memory controller. We found that this noise could be safely ignored.
 
\begin{table}
\centering
\caption{\new{Approximate total execution time for the experiments. $d$, $m$ and $s$ stand for \emph{day(s)}, \emph{minute(s)} and \emph{second(s)}.}}
\label{tab:exectime}
{
\resizebox{0.75\columnwidth}{!}
{
\begin{tabular}{c|cccccc}
\hline
 & \multicolumn{6}{c}{\textbf{Execution time}}                                                                                                                   \\ \hline
 & \multicolumn{1}{c|}{Eviction} & \multicolumn{5}{c}{\cellcolor{Gray}$62m$}                                                    \\ \cline{2-7} 
 & \multicolumn{1}{c|}{}         & \multicolumn{1}{c|}{E1} & \multicolumn{1}{c|}{E2} & \multicolumn{1}{c|}{E3} & \multicolumn{1}{c|}{E4} & E5  \\ \cline{3-7} 
 &
  \multicolumn{1}{c|}{\multirow{-2}{*}{{Previction}}} &
  \multicolumn{1}{c|}{\cellcolor{Gray}$0.7d$} &
  \multicolumn{1}{c|}{\cellcolor{Gray}$6s$} &
  \multicolumn{1}{c|}{\cellcolor{Gray}$7m$} &
  \multicolumn{1}{c|}{\cellcolor{Gray}$1.5d$} &
  \cellcolor{Gray}$> 1s$ \\ \cline{2-7} 
 & \multicolumn{1}{c|}{}                           & \multicolumn{1}{c|}{E6} & \multicolumn{1}{c|}{E7} & \multicolumn{1}{c|}{E8} & \multicolumn{1}{c|}{E9} & E10 \\ \cline{3-7} 
\multirow{-5}{*}{\rotatebox[origin=c]{90}{\textbf{Case study}}} &
  \multicolumn{1}{c|}{\multirow{-2}{*}{{Prefetching}}} &
  \multicolumn{1}{c|}{\cellcolor{Gray}$4d$} &
  \multicolumn{1}{c|}{\cellcolor{Gray}$6m$} &
  \multicolumn{1}{c|}{\cellcolor{Gray}$> 1s$} &
  \multicolumn{1}{c|}{\cellcolor{Gray}$> 1s$} &
  \cellcolor{Gray}$0.2d$ \\ \hline
\end{tabular}}
}
\vspace{-1em}
\end{table}

\section{Application}
\label{sec:experiments}
We show the utility of \LT{}s and the effectiveness of \framework through three case studies: cache eviction policy, previction and prefetching.
For each case the presented \LT{} is only a fragment of the identified \LT{}.
Missing cases are omitted due to either (1) clarity, i.e., complex relations are omitted, or (2) inconclusive outcome,
i.e., cases that generate random behavior and may not be reliably exploited.
Experiments are done on Raspberry Pi 3 \& 4--- widely
 used ARMv8 platforms, which use Cortex-A53 and -A72 CPUs, respectively.
On these processors, data is transferred between memory and cache in blocks of $64$~bytes and the L1 data cache is 4-way set associative.

Cortex-A53 cores show the previction behavior, and although the CPU is well-documented, the \emph{exact} behavior of the prefetcher on these cores is unknown.
For example, the ARM manuals do not answer any of the following questions: 
(\textbf{Q1})~How much data is prefetched?
(\textbf{Q2})~Do non-memory operations influence the prefetching behavior? And
(\textbf{Q3}), do load operations in one page affect prefetching due to memory loads from a different page? 

\new{To perform the experiments, we used a cluster of five Raspberry Pi~3 and one Pi~4 boards. %
\autoref{tab:exectime} shows the approximate total execution time to perform each experiment. Note that each experiment is a one-time effort required once per side channel and architecture.}

\subsection{Case Study: Eviction}
\label{sec:eviction}
To best illustrate the utility of \LT{}s, we use \framework to analyze the ARM Cortex-A72 cache eviction behavior. For experiment generation we leverage the parameterized eviction strategy from Rowhammer.js~\cite{DBLP:conf/dimva/GrussMM16} for x86 architectures and adopted in ARMageddon~\cite{DBLP:conf/uss/LippGSMM16} for ARM CPUs. The eviction strategy is shown in \autoref{alg:evict}.

\begin{algorithm}
\DontPrintSemicolon
   \For{$(s = 0; s \leq S - D ; s \pluseq L)$}
   {
  		\For{$(c = 0; c \leq C; c \pluseq 1)$}
  		{
             \For{$(d = 0; d \leq D; d \pluseq 1)$}
             {
                $*a[ s+d ];$
             }
         }
   }
\caption{Parameterized eviction strategy}\label{alg:evict}
\end{algorithm}

\begin{figure*}
	\centering
	\includegraphics[width=1\linewidth]{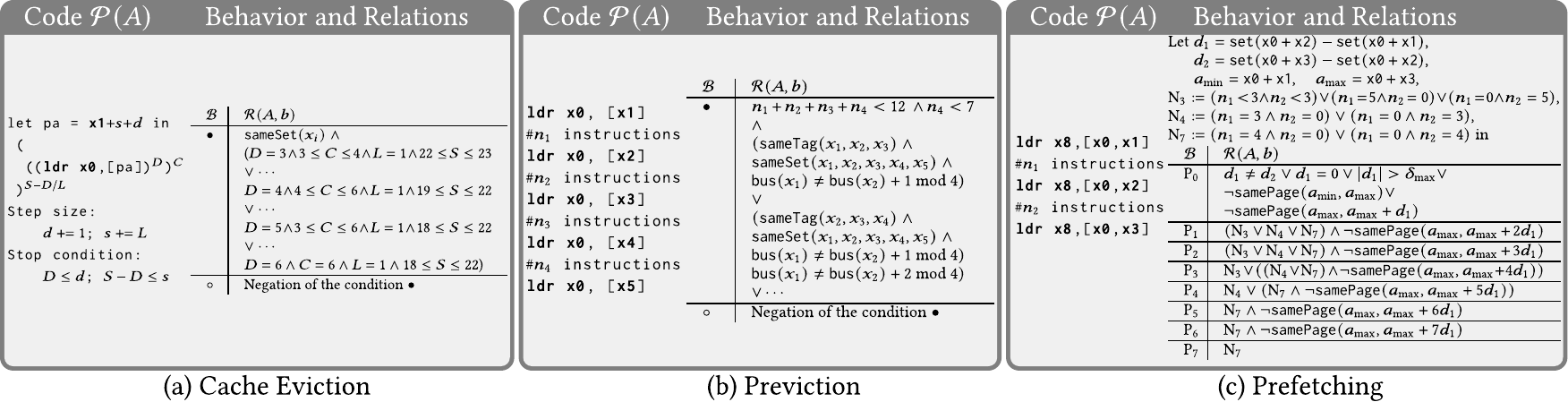}
\caption{Case studies' \LT{s} with selected relations. In (a) $a^b$ means $b$ times inlining repetition of instruction $a$. In (b), $\#{n_i}$ is inlining $n_i$ simple arithmetic, logical or \NOP instructions. For (a) and (b) triggering  and not triggering  behavior are denoted by $\bullet$ and $\circ$, respectively. In (c), $\text{P}_l$ denotes distinguishing behaviors and $l$ is the number of prefetched lines. Relations must be checked in order, the first matching relation determines the number of expected prefetches.
}
\Description{Three Leakage Templates.}
\label{fig:lts} 
\end{figure*}

For this experiment, we created the following \requestShort{} which corresponds to the parameterized eviction strategy:

$\rep{\pre{\memtwo{\tagtwo}{\setone} \s\expan{\expan{\expan{\memtwo{\tagone}{\setone}}{\memone{\mytag}, D, 1}}{C}}{\memone{\mytag}, S, L}}\memtwo{\tagtwo}{\setone}}{1000}$\\
where each experiment is repeated $1000$ times. 
The results were classified based on the existence of the pre-loaded address, indicating whether eviction occurred.
Based on the results, the \LT{} in \autoref{fig:lts}.a was generated. The generated \LT{} has an error tolerance of $5\%$, i.e., testcases that lead to eviction in more than $95\%$ of the times where classified as triggering ($\bullet$) behaviors.

The eviction test characterized by the parameterized eviction strategy is not a typical use case for \framework. However, it is suitable for illustrating \LT{}s. This test also demonstrates how \framework facilitates reverse engineering tasks as will be shown in \autoref{sec:discussion}.    

\subsection{Case Study: Previction}
\label{sec:previction}
We used \tool{}~\cite{scamv} to generate a set of \counterexample{es}, i.e., programs that may cause previction on the ARM Cortex-A53 (see~\autoref{sec:scamv}). 
All the generated examples were formed of exactly five load instructions. 
Moreover, looking at the cache content, it was evident that all \counterexample{es} loaded three different tags to the same cache set (and optionally a fourth tag to a different set), i.e., two to three loads targeted the same cache line.
We exploit this knowledge to construct initial \requestShort{es}.
We use these \requestShort{es} in five experiments (E1--E5) to iteratively refine an \LT{} for previction:

\parhead{E1: Minimal Code.}
\label{exp1}
First, we wanted to check whether the obtained \counterexample{es} contain minimal programs that trigger previction. 
We created a \requestShort{} which generates testcases containing all possible subsets (without repetition) of all instructions of each \counterexample{}. The \requestShort\ further mutated the word offsets for each of the loaded addresses. 
None of the generated testcases triggered previction. Thus, the \counterexample{es} contained minimal previction programs.

Performing \textbf{E1} requires doing 4 sets of experiments, each consisting of $5!/(5-x)! \times 16^x$ testcases, where $1\leq x \leq 4$. \autoref{tab:exectime} shows the total execution time to perform this experiment.

\parhead{E2: Order of Instructions.}
Next, we checked whether the order of instructions affects previction.
We created \requestShort{es} which generate testcases containing all possible permutations of instructions in distinct \counterexample{es}, e.g.,
\shuf{\memtwo{\tagone}{\setone}\;\ \memtwo{\tagone}{\setone}\ \memtwo{\tagone}{\setone}\ \memtwo{\tagtwo}{\setone}\ \memtwo{\tagthree}{\setone}}. In this GTS, the first three loads (denoted as \underline{1}, \underline{2} and \underline{3}) target the same cache line, but at different offsets.
The outcome of the permutation is shown in \autoref{tbl:order}.
\ce (first row) denotes previction and \nce (second row) denotes that no previction occurred.
We draw three conclusions from this test:
\begin{itemize}
\item{\emph{Relation on tags:}}
The relation between the location of tags affects previction.
Only programs with three consecutive load instructions with the same tag cause previction.
\item{\emph{Tag value:}} 
The exact tag value and location of the three consecutive loads does not matter, and can be similarly arbitrary for the non-consecutive load instructions.
\item{\emph{Word offset:}} 
Previction behavior differs based on the byte (word) offset of the loaded addresses.
In our example, multiple programs with the same order of cache sets and tags are in both \ce and \nce, e.g., $2$-$1$-$3$-$5$-$4$ (\ce) and $2$-$3$-$1$-$5$-$4$ (\nce). 
These permutations have instructions $1$ and $3$ swapped, i.e., two load instructions that only differ in their word offset.
\end{itemize}

\begin{table}
\centering
\def\arraystretch{1}
	\caption{Example permutation outcome. Each number represents an instruction from the initial testcase. Underlined numbers are loads from addresses that have the same tag.}
   \label{tbl:order}
	\resizebox{0.8\columnwidth}{!}{
		\begin{tabular}{ l | c }
			\hline 			
			\multirow{3}{*}{\bf	\ce} & \underline{\color{black}1}-\underline{\color{black}2}-\underline{\color{black}3}-{\color{gray}4}-{\color{gray}5}, \underline{\color{black}1}-\underline{\color{black}2}-\underline{\color{black}3}-{\color{gray}5}-{\color{gray}4}, \underline{\color{black}2}-\underline{\color{black}1}-\underline{\color{black}3}-{\color{gray}4}-{\color{gray}5}, \underline{\color{black}2}-\underline{\color{black}1}-\underline{\color{black}3}-{\color{gray}5}-{\color{gray}4}, \\
												& {\color{gray}4}-\underline{\color{black}1}-\underline{\color{black}2}-\underline{\color{black}3}-{\color{gray}5}, {\color{gray}4}-\underline{\color{black}2}-\underline{\color{black}1}-\underline{\color{black}3}-{\color{gray}5}, {\color{gray}4}-\underline{\color{black}3}-\underline{\color{black}1}-\underline{\color{black}2}-{\color{gray}5}, {\color{gray}4}-\underline{\color{black}1}-\underline{\color{black}3}-\underline{\color{black}2}-{\color{gray}5}, \\	
												& {\color{gray}5}-\underline{\color{black}1}-\underline{\color{black}2}-\underline{\color{black}3}-{\color{gray}4}, {\color{gray}5}-\underline{\color{black}2}-\underline{\color{black}1}-\underline{\color{black}3}-{\color{gray}4}, {\color{gray}5}-\underline{\color{black}3}-\underline{\color{black}1}-\underline{\color{black}2}-{\color{gray}4}, {\color{gray}5}-\underline{\color{black}1}-\underline{\color{black}3}-\underline{\color{black}2}-{\color{gray}4} \\	
			\hline					\hline		
  		     \bf \nce  & \underline{\color{black}2}-\underline{\color{black}3}-\underline{\color{black}1}-{\color{gray}5}-{\color{gray}4}, $\cdots,$ \underline{\color{black}2}-\underline{\color{black}3}-\underline{\color{black}1}-{\color{gray}4}-{\color{gray}5}, $\cdots,$ \underline{\color{black}3}-{\color{gray}4}-\underline{\color{black}1}-{\color{gray}5}-\underline{\color{black}2}, $\cdots$ \\ 
	\hline 			
		\end{tabular}}
\end{table}

\parhead{E3: Tags and Sets.}
We also checked the effect of \emph{exact tag and set values} on previction.
We created \requestShort{es} which generate testcases that preserve the relations between tags and/or sets of loaded addresses, while randomly changing these addresses:

(1)~\rep{\expan{\memtwo{\tagone}{\setone}}{3}\s\memtwo{\tagtwo}{\setone}\s\memtwo{\tagthree}{\setone}}{10000}

(2)~\rep{\memtwo{\tagtwo}{\setone}\s\expan{\memtwo{\tagone}{\setone}}{3}\s\memtwo{\tagthree}{\setone}}{10000}

(3)~\rep{\memtwo{\tagtwo}{\setone}\s\expan{\memtwo{\tagone}{\setone}}{2}\s\memtwo{\tagfour}{\settwo}\s\memtwo{\tagthree}{\setone}}{10000}

For every \requestShort, all $10{,}000$ generated testcases show the same behavior. Thus, the exact values of tags and sets do not matter.

\parhead{E4: Word Offset Behavior.}
In E2, we observed that the byte offsets of loaded addresses affect previction.
To broaden our understanding, in this experiment, we leveraged \requestShort{es} as shown in \autoref{tbl:requests}.
They generate testcases for 5-load programs with all possible combinations of tags and sets (for loads targeting up to two cache sets) while \new{mutating} the word offset. 
On Cortex-A53, each cache line ($64$~bytes) is divided into four disjoint ``buses'' of $16$~bytes (i.e. a cache line is loaded in $4$ bus rounds).
For example, for three tag-identical loads in a sequence of five set-identical loads, previction occurs if the bus of the first load is not the direct successor to the bus targeted by the second load.

This experiment took approximately 1.5 days to complete. The most time consuming part of this experiment is the case number 12 in~\autoref{tbl:requests} which consists of $12\times16^5$ testcases.

\parhead{E5: Priming the Cache.}
We further checked if previction also affects data cached before the execution of a testcase. We created \requestShort{es} which generate and execute previction testcases while $1$-$4$ lines of a cache set are occupied:

(1)~\pre{\memtwo{\tagfour}{\setone}}\expan{\memtwo{\tagone}{\setone}}{3}\s\memtwo{\tagtwo}{\setone}\s\memtwo{\tagthree}{\setone}

(2)~\pre{\memtwo{\tagfour}{\setone}\s\memtwo{\tagfive}{\setone}}\expan{\memtwo{\tagone}{\setone}}{3}\s\memtwo{\tagtwo}{\setone}\s\memtwo{\tagthree}{\setone}

(3)~\pre{\memtwo{\tagfour}{\setone}\s\memtwo{\tagfive}{\setone}\s\memtwo{\tagsix}{\setone}}\expan{\memtwo{\tagone}{\setone}}{3}\s\memtwo{\tagtwo}{\setone}\s\memtwo{\tagthree}{\setone}

(4)~\pre{\memtwo{\tagfour}{\setone}\s\memtwo{\tagfive}{\setone}\s\memtwo{\tagsix}{\setone}\s\memtwo{\tagseven}{\setone}}\expan{\memtwo{\tagone}{\setone}}{3}\s\memtwo{\tagtwo}{\setone}\s\memtwo{\tagthree}{\setone}

The results show that when the targeted set is half full (the second \requestShort), one of the two preloaded cache lines is evicted.
As we will show, this insight may result in a side-channel attack (see~\autoref{sec:attacks}).

\parhead{Previction Leakage Template.}
Our experiments resulted in \LT{}s that allow us to identify previction side channels in applications. 
\autoref{fig:lts}.b illustrates an example \LT{}.
Each of the experiments substantially refined this template.
\textbf{E1} dictated the general structure of five loads; \textbf{E2} contributed to the set and tag affinity; \textbf{E3} showed that we do not need to constrain certain tag/set values; \textbf{E4} revealed the bus relationship; and \textbf{E5} gave auxiliary information about previction behavior when caches are primed.
\textbf{E7} explains how we derived the wildcard instructions between the loads. 
\begin{table}
	\centering
	\def\arraystretch{1}
	\caption{Example \requestShort{es} used as input to \framework}
   \label{tbl:requests}
	\resizebox{0.8\columnwidth}{!}{
		\begin{tabular}{ c l c }
			\hline
				&		 \bf	Requests & \bf description  \\	
			\hline	
(1)		&		\fuzz{{\expan{\mem}{5}}}	& $1$ \ltag\xspace  \& $1$ \lset	 \\ 
	\hline
(2)		&		\fuzz{{\expan{\memtwo{\tagone}{\setone}}{4}\s\memtwo{\tagtwo}{\setone}}}	& \multirow{2}{*}{$2$ \ltags\xspace  \& $1$ \lset}	 \\ 
	
(3)		&		\fuzz{{\expan{\memtwo{\tagone}{\setone}}{3}\s\expan{\memtwo{\tagtwo}{\setone}}{2}}}	\\ 
	\hline
		&			\s\s\s\s\s\s\s\s\s\s\s\s\s\s$\cdots$ &			\\
		\hline
(12)		&		\fuzz{{\memtwo{\tagone}{\setone}\s\memtwo{\tagtwo}{\setone}\s\memtwo{\tagthree}{\setone}\s\memtwo{\tagfour}{\setone}\s\memtwo{\tagfive}{\settwo}}}	& $5$ \ltags\xspace  \& $2$ \lsets	 \\ 
\hline
		\end{tabular}}
\end{table}

\subsection{Case Study: Prefetching}
\label{sec:prefetching}
Based on the ARM Cortex-A53's reference manual, prefetching could leak the
(1)~number of loads, 
(2)~relation between loaded addresses (a.k.a.~\emph{stride}),
(3)~cache miss occurrences, and
(4)~end of a page.
These characteristics are not sensitive, as they can be extracted from the cache content even in the absence of prefetching.
We aim at validating the documented behavior, and also examine whether the undocumented behavior could leak sensitive information.
To this end, we again design five experiments (E6--E10).

\parhead{E6: \new{Prerequisites} for Prefetching.}
First, we wanted to devise necessary conditions for prefetching and determine the number of prefetched cache lines (\textbf{Q1}).
We created \requestShort{es} which generate testcases for all possible programs consisting of (\new{three to five})\footnote{\new{We chose this range as 1--2 loads do not trigger prefetching, and more than 5 loads would create too many testcases.}} loads while \new{mutating} their set index:

\cfuzz{\expan{\memtwo{\tagone}{\setone}}{3}},\;\; \cfuzz{\expan{\memtwo{\tagone}{\setone}}{4}},\;\; \cfuzz{\expan{\memtwo{\tagone}{\setone}}{5}}.\\
The main outcome of this test is an \LT{} describing the relations between loaded cache lines and the number of prefetched addresses.
Consider the \requestShort{} \memtwo{\tagone}{\setone}\s\memtwo{\tagone}{\settwo}\s\memtwo{\tagone}{\setthree}. Prefetching occurs when
$\setthree - \settwo = \settwo - \setone \leq \delta_{max}$, where $\delta_{max}=4$ denotes the maximum stride.
Moreover, programs with $3$--$4$ consecutive loads trigger the prefetching of $3$ additional cache lines, while streams with $5$ loads lead to $4$ prefetched addresses.

For prefetching, \textbf{E6} is the most time consuming case, as it consists of 3 sets of experiment each containing $2^{(7\times x)}$ testcases of $x$-loads programs, where $x\in\{3,4,5\}$. For the $x=3$ case we have done the full experiment but for the two other cases we have fixed a few bits (1 and 2 bits resp.) of the set indices to make the experiment manageable. Overall this experiment took 4 days to complete: 0.2d for $x=3$, 1d for $x=4$ and 2.8d for $x=5$.

\parhead{E7: Intermediate Instructions.}
Next, we wanted to check the effect of intervening instructions on prefetching (\textbf{Q2}).
For this, we have created \requestShort{es} which generate testcases containing programs with a fixed stride and a varying number of intermediate arithmetic instructions. For $0<n\leq 100$ and $0<m\leq 30$ we created:

(1)~\memtwo{\tagone}{\setone}\s\expan{\arith}{n}\s\memtwo{\tagone}{\setone+1}\s\memtwo{\tagtwo}{\setone+2}\s\memtwo{\tagone}{\setone+3}

(2)~\memtwo{\tagone}{\setone}\s\memtwo{\tagone}{\setone+1}\s\expan{\arith}{n}\s\memtwo{\tagtwo}{\setone+2}\s\memtwo{\tagone}{\setone+3}

(3)~\memtwo{\tagone}{\setone}\s\memtwo{\tagone}{\setone+1}\s\memtwo{\tagtwo}{\setone+2}\s\expan{\arith}{n}\s\memtwo{\tagone}{\setone+3}

(4)~\memtwo{\tagone}{\setone}\s\expan{\arith}{m}\s\memtwo{\tagone}{\setone+1}\s\expan{\arith}{m}\s\memtwo{\tagtwo}{\setone+2}\s\expan{\arith}{m}\s\memtwo{\tagone}{\setone+3}\\
We created similar \requestShort{es} for 3- and 6-load streams.

Our results showed that adding instructions between consecutive loads could alter the number of prefetched addresses, e.g., adding $3$ arithmetic instructions between two consecutive loads can increase the number of prefetched addresses from $3$ to $4$. 
Similarly, adding $4$ arithmetic instructions increases this number to $7$ and adding $5$ instructions reduces it again to $3$. Thus, the prefetcher may leak the control flow at the granularity of one instruction, a new insight which may lead to potential side channels (see~\autoref{sec:attacks}).

\parhead{E8: Respecting Page Boundary.}
We also checked whether prefetching respects page boundaries (as stated in the manual), i.e., if the processor prefetches addresses past the end of a page (\textbf{Q3}). To this end, we created \requestShort{es} to generate testcases containing programs with fixed strides while gradually shifting the loaded addresses toward the next page, i.e., up to one page ($64*64=4096$). For $0<n\leq 5$ we created:

(1)~\slide{\memtwo{\tagone}{\setone}\s\memtwo{\tagone}{\setone+n}\s\memtwo{\tagone}{\setone+2n}}{64}

(2)~\slide{\memtwo{\tagone}{\setone}\s\memtwo{\tagone}{\setone+n}\s\memtwo{\tagone}{\setone+2n}\s\memtwo{\tagone}{\setone+3n}}{64}

(3)~\slide{\memtwo{\tagone}{\setone}\s\memtwo{\tagone}{\setone+n}\s\memtwo{\tagone}{\setone+2n}\s\memtwo{\tagone}{\setone+3n}\s\memtwo{\tagone}{\setone+4n}}{64}\\
The results show that testcases at the end of the page had fewer prefetched cache lines as not to cross page boundary. Testcases with loads spread across different pages did not cause prefetching.

\parhead{E9: Multiple Prefetching Sequences.}
We now explore how the prefetcher handles multiple, possibly interleaving sequences.
To this end, we specify a GTS that merges three 3-load sequences with distinct tags and sets, i.e., from different memory pages.

 {\fontsize{9.3}{9.8}{\mulmerge{\merge{\memtwo{\tagone}{\setone}\;\memtwo{\tagone}{\setone+1}\;\memtwo{\tagone}{\setone+2}}{\memtwo{\tagtwo}{\settwo}\s\memtwo{\tagtwo}{\settwo+1}\s\memtwo{\tagtwo}{\settwo+2}}}{:\memtwo{\tagthree}{\setthree}\s\memtwo{\tagthree}{\setthree+1}\s\memtwo{\tagthree}{\setthree+2}}}}\\
Our results show that the prefetcher becomes active for only the first two sequences; any additional sequence will \emph{not} be prefetched.
To decide which one is first, the prefetcher picks the first sequences of three consecutive loads (two strides).
This means that multiple independent sequences can cause interference, even if they are on different pages (\textbf{Q3}).
\new{Again, this novel observation can lead to potential side channels to leak information (\autoref{sec:attacks}).}

\parhead{E10: Cache Hits.}
Finally, we tested the influence of cache hits on prefetching. 
We created \requestShort es, which generated and executed prefetching testcases while one of the loaded addresses is cached:

(1)~\pre{\memtwo{\tagone}{\setone}}\cfuzz{\memtwo{\tagone}{\setone}\s\memtwo{\tagone}{\setone+1}\s\memtwo{\tagone}{\setone+2}} 

(2)~\pre{\memtwo{\tagtwo}{\settwo}}\memtwo{\tagone}{\setone}\s\memtwo{\tagtwo}{\settwo}\s\memtwo{\tagone}{\setone+1}\s\memtwo{\tagone}{\setone+2}

The GTS (1) did not trigger prefetching for all generated testcases, while the GTS (2) induced a behavior similar to that of \memtwo{\tagone}{\setone} \ \memtwo{\tagone}{\setone+1}\s\memtwo{\tagone}{\setone+2}.
Thus, the prefetcher only monitors cache misses, i.e., preloaded data may  destroy sequences that would have otherwise been prefetched.
This is problematic if prefetched sequences from different pages (and contexts) interfere with each other.

\parhead{Prefetching Leakage Template.}
\autoref{fig:lts}.c illustrates an \LT{} for a 3-load stream, allowing to identify prefetching side channels in applications. 
To construct this \LT:
\textbf{E6} identified constraints on the cache sets;
\textbf{E7} derived bounds on intermediate instructions and their effect on the prefetching behavior;
\textbf{E8} refined constraints on cache sets based on page boundaries;
\textbf{E9} revealed the interference between interleaving prefetching sequences; and
\textbf{E10} gave auxiliary information about the effect of cache hits on  prefetching. 
\section{Matching Prefetching \LT{} in Binaries}
\label{sec:reidentifyingPoC}
We now demonstrate how an \LT{} can be used to identify an instance of the side channel it describes in a target binary.

\parhead{Target Binary and Side Channel.}
As a proof of concept, we re-identify a known prefetching-based side channel in OpenSSL 1.1.0g. It was first found and exploited on an Intel CPU by Shin \etal{} \cite{10.1145/3243734.3243736}. 
Data-dependent loads from a lookup table may or may not trigger the prefetcher to load certain cache lines into the cache, depending on the resulting memory access pattern. Therefore, the cache state of potentially prefetched cache lines indicates the existence of relations between the accessed lookup table elements and, by extension, the processed data. Shin \etal{} exploit these relations to leak the scalar of a scalar point multiplication on an elliptic curve. In Elliptic Curve Diffie-Hellman (ECDH), a scalar represents the private key. The attack recovers the key incrementally. The same computation is applied to both the target scalar and a candidate scalar. By changing the candidate scalar such that the prefetching behavior assimilates, both scalars assimilate as well.
Even though this vulnerability is no longer present in recent OpenSSL versions, we still consider it a reasonable case study to demonstrate that \LT{s} can be used to identify real-world vulnerabilities in binaries.

\parhead{Approach: Combining Static and Dynamic Analysis.}
Shin \etal{} \cite{10.1145/3243734.3243736} limit the scope of their search to a specific cryptographic operation. In contrast, our starting point is the whole OpenSSL binary.
We combine static and dynamic binary analysis techniques to search it for instances of the prefetching \LT{} (see \autoref{fig:lts}.c).
First, we scan the binary for code sections that match the code pattern $\proghat$ of the \LT{}. This results in a list of candidate code sections that potentially contain a prefetching side-channel.
Second, we need to check whether a candidate section satisfies different relations $\relation$ for different input values. If this is the case, we expect the section to show input-dependent behavior, indicating a side channel.
Not all relations can be resolved statically, especially if they refer to addresses in instruction operands. To overcome this, we dynamically analyze the target code to learn its concrete addresses.

\parhead{Performing Static Analysis.}
We use \emph{asmregex} \cite{asmregex} to statically scan the target binary for the code pattern $\proghat$ of the prefetching \LT{}. Asmregex searches binaries for code sections that match a specified pattern. We extended the tool by approx. 200 LoC (code available at \cite{plumber}) to support a subset of the ARM instruction set and added support for backreferences to the pattern language. Backreferences allow to express simple relations between instructions. For instance, two subsequent load instructions can be required to use the same base address register.
To identify code sections matching $\proghat$ in OpenSSL, we convert $\proghat$ into an asmregex pattern.
This pattern matches 429 3-load sequences across 18 OpenSSL modules. By briefly inspecting the matching candidate sections, we identify accesses to lookup tables in 11 of these modules.
The remaining matches are predominantly caused by operations on complex data structures.
Most importantly, we identify the code section exploited in \cite{10.1145/3243734.3243736} among the candidates in the module \texttt{crypto/bn/bn\_gf2m.o}. Further investigation of other matches is considered out of scope.

\parhead{Performing Dynamic Analysis.}
We proceed with the dynamic analysis step to check a candidate code section for input-dependent behavior using the relations $\relation$ from the \LT{}. We create a simple wrapper program that calls the matching library function with varying input values. This program can be used to log all (input-dependent) loads from the relevant lookup table \texttt{SQR\_tb}, which spans across three cache lines in memory. 
We record two different traces for each input value.
First, we use Valgrind \cite{valgrind} and GDB \cite{gdb} to record an \emph{access trace}, a list of all loads from \texttt{SQR\_tb} during program execution. This trace can be used to determine the \emph{expected} prefetching behavior based on the relations $\relation$ from the \LT{}.
Second, we use a Flush+Reload side channel to record a \emph{cache trace}. This trace contains the cache state of the memory lines around \texttt{SQR\_tb} after execution. It is captured for evaluation purposes and indicates the \emph{actual} prefetching behavior of the CPU.

In order to show that the \LT{} accurately represents the prefetching behavior, we recorded traces for 100 random input values to the library function. For each input value, we determined the expected prefetching behavior using the access trace\footnote{\new{As we found in \autoref{sec:prefetching}-\textbf{E10} that the prefetcher only operates on cache misses, the load instructions relevant to the prefetcher are not necessarily the first three load instructions in the matching code section. Therefore, we perform our analysis based on the first three loads in each access trace that target different cache lines.}} and compared it with the actual behavior using the corresponding cache trace.

\parhead{Evaluation.}
\autoref{tab:matching_confusion} illustrates the classification performance. For all 66 cases where the load instructions satisfy the relations for $\text{P}_0$, the cache traces show that no prefetching occurred.
In six cases, the relations for behavior $\text{P}_3$ are satisfied. The three relevant load instructions load data from three consecutive cache lines and the number of instructions between the load instructions ($n_1$ and $n_2$) is within the specified bounds. In all six cases, the cache trace shows that prefetching of three additional cache lines occurred.
In the remaining 28 cases, the relations for none of the behaviors from the \LT{} are satisfied. The reason is that the distances $n_1$ and $n_2$ between the relevant load instructions are outside the parameter range we tested when the \LT{} was created. We denote these cases as \emph{undecidable} cases.
We note that no misclassifications occurred.

\parhead{Conclusion.}
We successfully demonstrated that the prefetcher of the Cortex-A53 CPU shows input-dependent behavior for the library function under investigation. This is the base requirement for the differential attack in \cite{10.1145/3243734.3243736}. The \LT{} helped us to re-identify this vulnerability known from the Intel architecture in ARM binary code. In contrast to prior work, our starting point was the whole OpenSSL code base.
For code sections that closely match the \LT{} (i.e., they closely correspond to code and relations that \framework{} encountered during creation of the \LT{}), the behavior classification based on the relations is accurate. When unknown relations occur, undecidable cases are more likely to appear. In our example, undecidable cases occur due to higher values of $n_1$ and $n_2$ than we used when creating the \LT{} (to keep the number of test cases within a reasonable range). However, these cases can be detected and the analyst may use them to design further experiments in order to refine the \LT{} in a targeted manner. This highlights again that a \LT{}, which can hardly be ever complete, can be developed in an iterative fashion.

\begin{table}
	\caption{Confusion matrix, comparing prefetching behavior classification based on relations with the actual behavior.}
	\label{tab:matching_confusion}
	\resizebox{0.85\columnwidth}{!}
    {
	\begin{tabular}{cc|C{5em}|C{5em}|C{5.5em}}
		\multicolumn{2}{c}{}&\multicolumn{3}{c}{\textbf{Relation-based classification}}\\
		\cline{3-5}
		\multicolumn{2}{c}{} & \emph{\textbf{$\text{P}_0$}} & \emph{\textbf{$\text{P}_3$}} & \emph{\textbf{undecidable}}\\
		\cline{2-5}
		\multirow{2}{*}{\shortstack[r]{\textbf{Actual}\\\textbf{behavior}}} & \emph{\textbf{$\text{P}_0$}} & 66 &0 & 0\\
		\cline{2-5}
		& \emph{\textbf{$\text{P}_3$}} & 0 & 6 & 28\\
		\cline{2-5}
	\end{tabular}
	}
\end{table}

\section{Novel Leakage Primitives}
\label{sec:attacks}

\definecolor{Gray}{gray}{0.9}
\begin{table}
\caption{Transmission and error rates of sota. covert channels.}
\label{tab:covert-speed}
\resizebox{0.75\columnwidth}{!}
{
 \begin{tabular}{lrr}
  \toprule
  \textbf{Covert channel (Element)} & \textbf{Speed} & \textbf{Error rate} \\
  \midrule
  Liu~\etal~\cite{Liu2015Last} (L3) & {600} {\kilo\bit/\second} & {1} {\percent} \\
  Pessl~\etal~\cite{Pessl2016} (DRAM) & {411} {\kilo\bit/\second} & {4.11} {\percent} \\
  Maurice~\etal~\cite{Maurice2017Hello} (L3) & {362} {\kilo\bit/\second} & {0} {\percent} \\
  \rowcolor{Gray}
  \texttt{PRF\_IS} & {276} {\kilo\bit/\second} & {0.05} {\percent} \\
  \rowcolor{Gray}
  \texttt{PRF\_OS} & {206} {\kilo\bit/\second} & {2.1} {\percent} \\
  \rowcolor{Gray}
  \texttt{PRF\_CF} & {76} {\kilo\bit/\second} & {0.7} {\percent} \\
  \rowcolor{Gray}
  \texttt{PR\_FR} & {73} {\kilo\bit/\second} & {1.2} {\percent} \\
  Maurice~\etal~\cite{Maurice2015C5} (L3) & {751} {\bit/\second} & {5.7} {\percent} \\
  Wu~\etal~\cite{Wu2012} (memory bus) & {747} {\bit/\second} & {0.09} {\percent} \\
  Semal~\etal~\cite{Semal2020covert} (memory bus) & {480} {\bit/\second} & {5.46} {\percent} \\
  Schwarz~\etal~\cite{Schwarz2017Timers} (DRAM) & {11} {\bit/\second} & {0} {\percent} \\
  \bottomrule
 \end{tabular}
}
\end{table}

\begin{figure}
\centering
   \includegraphics[width=0.65\linewidth]{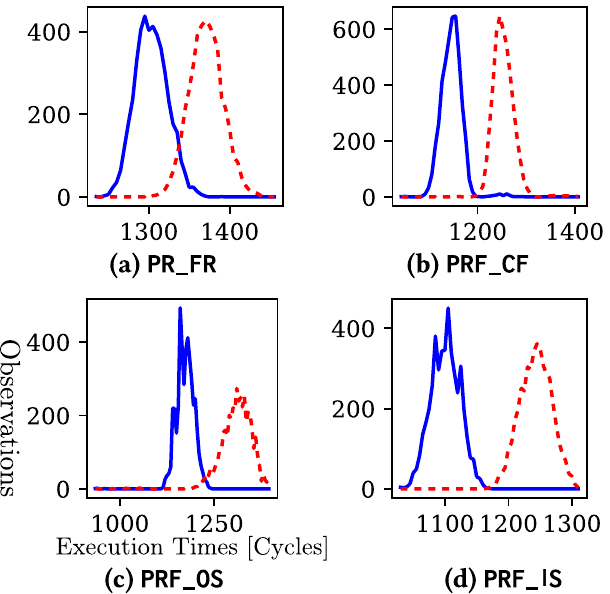}
   \caption{Histograms showing the execution time of the channel w.r.t.
   behavior  (solid blue line) or not (dashed red line).}
   \Description{The figure shows four histograms. The histograms are labeled "(a) PR\_FR", "(b) PRF\_CF", "(c) PRF\_OS", and "PRF\_IS". The x-axes are labeled "Execution Times [Cycles]", the y-axes are labeled "Observations". Each histogram contains two lines. A solid blue line represents the frequency distribution of execution times of the channel when the behavior is observed, the red line represents the execution time of the channel when the behavior is not observed. All lines form bell curves. In all four histograms, the peak of the blue bell curve is before (left to) the peak of the red bell curve, and the overlap of both curves is small. This indicates that both behaviors are clearly distinguishable. All execution times are in within the interval of [1000; 1500] cycles.}
   \label{fig:exectime_histograms}
\end{figure}

Our experiments in~\autoref{sec:experiments} also helped us to identify five novel prefetching based leakage primitives. 
For four of these primitives, we present a minimal code example and evaluation results of its leakage speed and error rate (see \autoref{tab:covert-speed} and \autoref{fig:exectime_histograms}). We omit the evaluation of \texttt{PR\_PP} as it is not applicable in our covert channel setup.

\begin{figure*}[t]
  \centering
\includegraphics[width=0.65\linewidth]{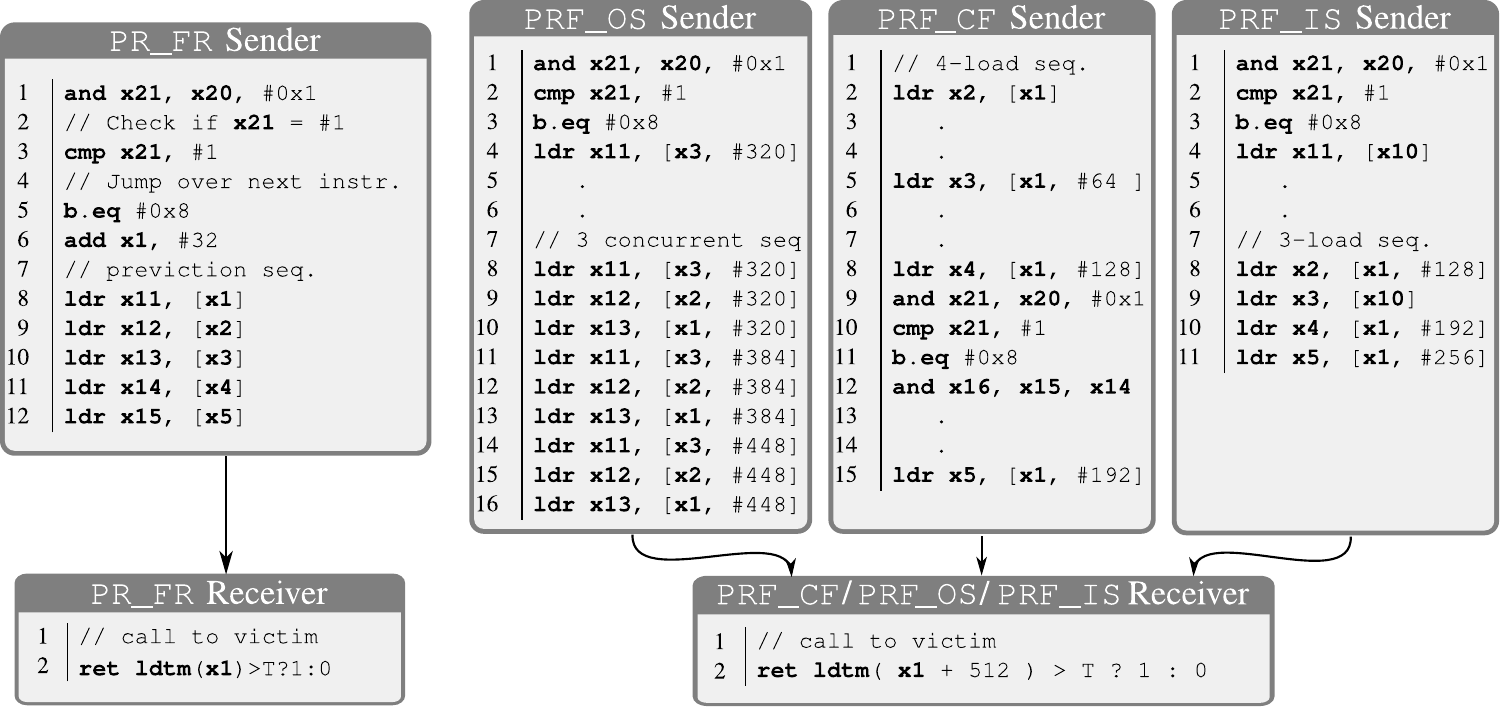}
\caption{Examples for leakage primitives. \textbf{\texttt{ldtm}} measures  the \lstinline|ldr|  execution time for the given address, \lstinline|T| is the threshold.}
\Description{The figure provides small assembly code snippets that implement sender and receiver for the leakage primitives described in section \ref{sec:attacks}.}
\label{fig:channels}
\end{figure*}

\subsection{Previction w/ Shared Memory (\texttt{PR\_FR})}
Our previction-based Flush+Reload primitive \texttt{PR\_FR}  is based on our insights from \textbf{E4} in \autoref{sec:previction}. 
Unlike traditional Flush+Reload primitives, \texttt{PR\_FR}  allows leaking information based on a \emph{bus} rather than cache lines.
The idea is to make the (observable) occurrence of previction dependent on a secret bit (the leak target) by changing bus relations.

The primitive \texttt{PR\_FR} in~\autoref{fig:channels} leverages the strong bus dependency between the consecutive load instructions in a valid previction sequence.
Let the lines $8$--$12$ be a valid previction sequence following our \LT{} in \autoref{fig:lts}.b, i.e.,  $\text{\regname{x1}}$--$\text{\regname{x3}}$ being consecutive loads with a ``valid'' bus relation, and $\text{\regname{x4}}$/$\text{\regname{x5}}$ arbitrary other loads from the same set.
The  idea is to use a secret-dependent conditional change to the byte offset of the first load ($\text{\regname{x1}}$) to destroy previction (lines $1$ through $6$).
That is, the word offset of the first address loaded (in $\text{\regname{x1}}$) depends on the value of the least significant bit of the data stored in $\text{\regname{x20}}$.
According to the \LT{}, the relation between the offsets in the addresses in $\text{\regname{x1}}$ and $\text{\regname{x2}}$ determines if $\text{\regname{x1}}$ will be previcted from the cache.
Thus, by measuring the time required to load from the address stored in $\text{\regname{x1}}$, the receiver can learn whether this address resides in the cache and consequently leak the secret bit.

\subsection{Previction w/o Shared Memory (\texttt{PR\_PP})}
Based on experiment \textbf{E5} in~\autoref{sec:previction}, previction may target preloaded memory addresses and leak information in the absence of shared memory, e.g., through Prime+Probe.
The sender code of our pre\-vic\-tion\--based Prime+Probe primitive \texttt{PR\_PP} is similar to that of \texttt{PR\_FR}.
However, in \texttt{PR\_PP}, the receiver first loads two memory lines into the targeted cache set before the execution of the sender code. The receiver then probes the lines to determine the leaked bits.

\subsection{Prefetching Control-Flow Leakage (\texttt{PRF\_CF})}
\texttt{PRF\_CF} allows leaking the control flow of a program based on prefetching. It is based on the results of \textbf{E7} in \autoref{sec:prefetching}.
\autoref{fig:channels} shows an example code of \texttt{PRF\_CF}. 
The sender code has a 4-load prefetching sequence
with a fixed stride (lines $2$, $5$, $8$, and $15$). The loads are separated by a number of arithmetic instructions.
The instruction at line $12$ is conditionally executed depending on one bit of a secret that is stored in $\text{\regname{x20}}$
(lines $9$ through $12$).
According to \textbf{E7}, the number of executed instructions within a prefetching sequence affects the number of prefetched cache lines.
	By measuring the time required to reload a (possibly prefetched) address $\text{\regname{x1}+512}$, 
the receiver can determine whether an instruction was executed and consequently learn the secret bit.

\subsection{\new{Prefetching on an Interrupted Seq. (\texttt{PRF\_IS})}}
Inspired by \textbf{E7}, we tested the effect of intermediate memory operations on prefetching.
We observed that an intermediate load from a different page leads to prefetching of additional cache lines by a 3-load stream.
\texttt{PRF\_IS} is based on this outcome. It also allows leaking accesses to non-shared memory through shared pages.

As shown in \autoref{fig:channels}, the sender code contains a 3-load prefetching sequence with a fixed stride and an interleaving load from a different page, i.e., $[\text{\regname{x10}}]$ (lines $8$ through $11$). 
Since the processor ignores cache hits, the number of prefetched lines will depend on whether $\text{\regname{x10}}$ is cached.
Consequently, by measuring the time required to load from $\text{\regname{x1}+512}$, the receiver can determine whether $\text{\regname{x10}}$ has been cached and consequently learn the secret bit.

In contrast to the prefetching experiment in \autoref{sec:prefetching}, \texttt{PRF\_IS} checks how the prefetcher's behavior changes when it \emph{observes} interleaved loads from different pages (i.e., across page boundaries). Note that in \texttt{PRF\_IS} all \emph{predicted} addresses are within bounds of the same page (in accordance with \textbf{E7}). 

\subsection{Prefetching and Outstanding Seq. (\texttt{PRF\_OS})}
\texttt{PRF\_OS} exploits competing prefetching sequences to leak accesses to non-shared memory through shared pages. In other words, it allows leaking secrets through Flush+Reload even when secret-dependent memory accesses are from non-shared memory.
\texttt{PRF\_OS}  is based on the outcome of experiments \textbf{E9} and \textbf{E10} in \autoref{sec:prefetching}. 

As shown in \autoref{fig:channels}, the sender has three interleaving 3-load streams with fixed strides (lines $8$, $11$, $14$; lines $9$, $12$, $15$; and lines $10$, $13$, $16$).
These streams are preceded by a load from the address $\text{\regname{x3}}$, whose execution depends on one bit of a secret stored in $\text{\regname{x20}}$ (lines $1$ through $4$).
According to \textbf{E9}, memory addresses are prefetched for sequences whose strides are detected first.
Additionally, according to \textbf{E9}, the processor ignores cache hits when detecting prefetching sequences. Consequently,
depending on whether $\text{\regname{x3}}$ is cached, prefetching would be triggered for either the first or the second two prefetching sequences.
By measuring the time required to load from $\text{\regname{x1}+512}$, 
the receiver can determine whether prefetching occurred for the third stream and consequently learn the secret bit.

\section{Related Work}

\parhead{Reverse-engineering cache behavior.}
Prior work on reverse-engineering the cache behavior mainly focused on replacement policies.
In particular, existing approaches aim at reverse-engineering known permutation-based replacement policies~\cite{MeCRP, ReCRP}, e.g., FIFO and PLRU, as well as new adaptive policies~\cite{LeCRP, RRIP}.
Some approaches pursued ad-hoc means~\cite{MeCRP, ReCRP}, others relied on a novel register automata learning technique~\cite{LeCRP} or compared hardware output against software-simulated caches~\cite{SkCRP, CQCRP}.
However, these methods are either not practical~\cite{LeCRP}, or not general and do not guarantee correctness~\cite{CQCRP}.
We rather design a framework to better understand undocumented behavior of hardware features such as previction~\cite{scamv}, eviction policies and the cache prefetcher and specify their leakage templates.
\framework can also be used to facilitate reverse engineering of microarchitectural components like the branch predictor.

\parhead{Side-channel attacks.}
Over the past decades, researchers have devised various means to exfiltrate secrets from computing devices based on electromagnetic~\cite{magneticSC}, power-based~\cite{powerSC}, and timing-based~\cite{primeprobe, flushreload, flushflush, Tsunoo03cryptanalysisof} side channels.
Timing-based side channels can be exploited purely in software, thus also remotely.
Most timing attacks exploit timing differences introduced by processor caches. Cache-based attacks proposed in the literature include but are not limited to Prime+Probe~\cite{Tromer:2010:ECA:1713125.1713127, primeprobe}, Flush+Reload~\cite{flushreload}, Evict+Time~\cite{evicttime}, and their variants~\cite{flushflush, cachetemplate}.
Such attacks target implementations of cryptographic algorithms~\cite{cryptosidechannels, timeRSA, evicttime, flushflush}, and more generic attack vectors such as keylogging~\cite{cachetemplate}.
Closely related to previction are CacheOut~\cite{schaik2020cacheout} and RIDL~\cite{ridl} which use cache eviction to leak secret information. 
Similarly, CacheBleed~\cite{DBLP:journals/jce/YaromGH17} exploits cache bank conflicts to break, resp., RSA and AES.

\parhead{\framework vs. \tool{}}
\framework{'s} goal is to facilitate understanding  microarchitectural behavior, e.g., triggers and effects, via easily constructed queries. 
The approach of \tool{}~\cite{scamv} (and similar tools, e.g.,~\cite{DBLP:conf/ndss/GrasGKBR20,DBLP:journals/corr/abs-2106-03470}) is complementary to that of \framework{}. This relation is better expressed as a 
two-step approach: first, \tool{} or similar tools are used to detect  a possible channel by specifying the monitored component, e.g., the cache state, which often is dictated by the vulnerability discovery tool (e.g., port contention in~\cite{DBLP:conf/ndss/GrasGKBR20}, or execution time in~\cite{DBLP:journals/corr/abs-2106-03470}).
Whenever the tool discovers yet-unsupported types of side channels on  monitored components, \framework{} can be used to learn the correlation between (attacker-controlled) inputs and the channel (see~\autoref{fig:framework}). The challenging aspect of such an integration is the generalization of concrete code examples, which can be provided by the analyst.

\tool{} finds a channel by executing randomly generated program-input pairs. Using an SMT solver to generate inputs requires careful engineering of the queries sent to the solver, otherwise generated inputs and thus counterexamples will be too similar which makes them not suitable for statistical analysis. Conversely, \framework{} makes input generation more efficient by using GTS and allows learning the correlation between inputs and the channel.
\section{Discussion}
\label{sec:discussion}
We proceed to discuss the utility of \framework{} in other use cases and its limitations. In particular, we show that \framework can facilitate reverse engineering of microarchitectural features. 
We show this by applying \framework to reverse engineer the \new{structure} of the Cortex-A53's branch predictor.

\begin{figure}
	\centering
	\includegraphics[width=1\linewidth]{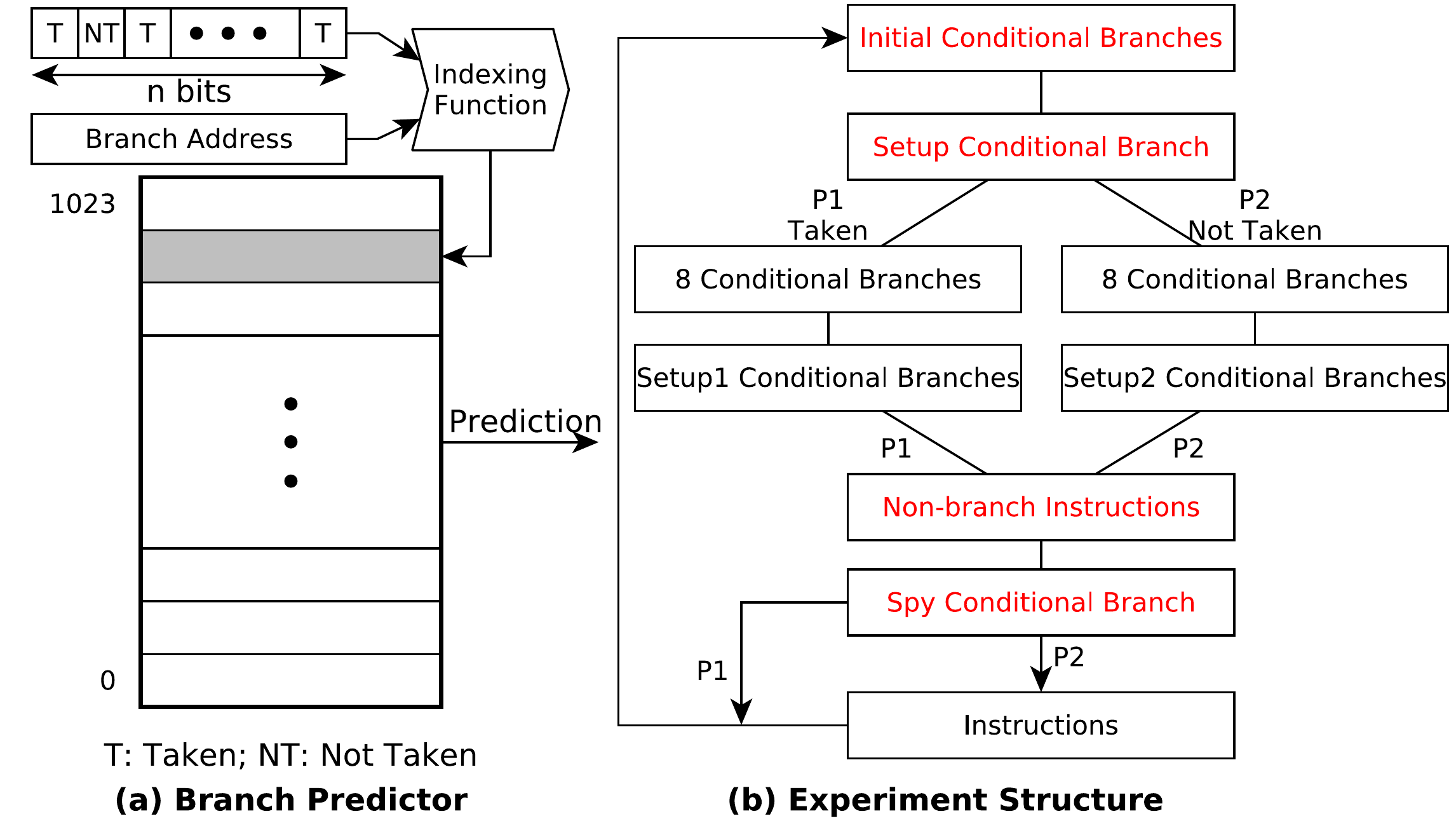}
	\caption{Branch predictor experiment.}
	\Description{The figure consists of two subfigures. The first subfigure is labeled "(a) Branch Predictor". It shows a Branch History Register (BHR) that contains n bits and is used as an index into a Pattern History Table (PHT). The second subfigure is labeled "(b) Experiment Structure". It shows a flowchart that represents the sequence of branch instructions that is used in the experiment code for the branch predictor experiment in section \ref{sec:discussion:bp}.}
	\label{fig:bpexperiment}
\end{figure}

\parhead{Reverse Engineering with \framework{}: Branch Predictor.}
\label{sec:discussion:bp}
The only available information on the Cortex-A53 branch predictor unit is that it is a \emph{global type} that uses branch history registers (BHRs) and a 3072-entry pattern history table (PHT)~\cite{cortexa53}. 
Each entry of the PHT contains a 2-bit saturating counter to predicate a branch outcome. The PHT is accessed via a BHR 
that stores the history of the recently executed branches (see Fig.~\ref{fig:bpexperiment}.a). 
The goal is to find the size and the number of BHRs, the structure of the PHT, and the mechanism used to map branches to this table. Our experiment is inspired by the work of Uzelac et al.~\cite{DBLP:conf/ispass/UzelacM09} to reverse engineer the branch predictor of Intel's Pentium M CPU.

Since the size of a pattern history table should be a power of $2$~\cite{DBLP:conf/sbac-pad/OzerRB07}, we assumed that Cortex-A53's branch predictor should be
structured as three tables with $1024$ entries, each connected to a unique BHR of size $10$~bits. 
Therefore, to ensure that every branch instruction is mapped to the same BHR and PHT, we additionally assumed that the branch address modulo $3$ is used to compute the index to access BHR and PHT.
Fig.~\ref{fig:bpexperiment}.b shows the general structure of the experiment (adapted from~\cite{DBLP:conf/ispass/UzelacM09}) we have conducted to validate our conjectures. In this experiment the \emph{initial conditional branches} are used
to force the branch predictor to mispredict the \emph{setup conditional branch}. \emph{Non-conditional instructions} are dummy instructions (e.g. \texttt{nop}) used to control the distance between the setup $1$/$2$ conditional branches and our \emph{spy conditional branch}. 
To ensure that all branches will be mapped to the same BHR and PHT we pad the distances between all effective branch instructions with \texttt{nop}s, i.e., making their address congruent modulo $3$.

To determine the size of the BHR and PHT we use a spy branch with controllable outcome pattern. The branch is always taken if reached through path $P1$ and not taken when reached through $P2$. Thus, when the number of BHR entries for P1 and P2 are equal, the spy branch will be mispredicted. Misprediction also happens if the spy branch's history is not present in the PHT. 
Since the spy branch is not the only mispredicted branch, we observe (utilizing the \textit{performance monitor unit} (PMU)) $\approx$100\% misprediction rate when the spy branch is mispredicted and $\approx$60\% otherwise. 

Uzelac et al.~\cite{DBLP:conf/ispass/UzelacM09} proposed an experiment to find the branch address bits which affect the BHR (when PHT is not full). In their experiment the observed misprediction rate is also either 60\% or 100\%.
Based on this experiment, we utilize \framework to discover the PHT size for Cortex-A53. To do so, we generate a \requestShort{} to fuzz the number of initial conditional branches and the distance between the spy branch and the setup 1 and 2 branches. The \requestShort{} is formed of the following blocks (see Fig.~\ref{fig:bpexperiment}.b):
\\
{
	\setlength{\tabcolsep}{2pt}
	\begin{tabular}{lr}
		\expan{\branchthree{\bidone}{\btrue}{12} \s\setbranchtwo{\bidtwo}{\btrue} \s\nop}{X} & (Initial conditional branches)\\

		\setbranchtwo{\bidone}{\bool} \s\branchthree{\bidone}{\bfalse}{53} & (Setup conditional branch)\\

		\expan{\branchthree{\bidtwo}{\btrue}{12} \s\setbranchtwo{\bidtwo}{\btrue} \s\nop}{8} \s\branchthree{\bidtwo}{\btrue}{12} \s\setbranchtwo{\bidtwo}{\btrue} & (Setup 1)\\

		\expan{\branchthree{\bidtwo}{\bfalse}{12} \s\setbranchtwo{\bidtwo}{\btrue} \s\nop}{8} \s\branchthree{\bidtwo}{\bfalse}{12} \s\setbranchtwo{\bidtwo}{\btrue} & (Setup 2)\\

		\expan{\nop}{Y} & (Non-branch instructions)\\

		\branchthree{\bidone}{\btrue}{12} \s\setbranchtwo{\bidtwo}{\btrue} & (Spy conditional branch)
	\end{tabular}
}
\\
For each tested combination of $X$ and $Y$, the concatenation of the above blocks is executed $10{,}240$ times, while alternating the value of \bool{} every $16$ executions. 
This is done by using the power macro as follows: \expan{\expan{\bool = \btrue}{16} \s\expan{\bool = \bfalse}{16}}{40}.

For any number $X$ of initial conditional branches less than $1024$ we got (almost) the same result as Uzelac (i.e. a misprediction rate between $60\%$ and $100\%$). However, when the PHT is full, the misprediction rate is always 100\%.
The results of these experiments support our conjecture on the size of PHT and BHR.

\new{\parhead{\framework{'s} limitations.}
Given the complexity of the microarchitectural components and the number of side-channel attacks, the current implementation of \framework{} mainly targets cache-based side channels and the implementation of some components is limited to ARM (ARM-v7 and -v8) and RISC-V architectures. However, \framework{'s} design is generic and not constrained to cache-related channels and its implementation can be ported to other architectures such as x86. 
The main challenge of such an adaptation is porting \framework{'s} inspection module, which is currently deployed in ARM TrustZone. Moving to a new platform, e.g. Intel or AMD, would require replacing the inspection module by an alternative probing mechanism such as Flush+Reload~\cite{flushreload} or Flush+Flush~\cite{flushflush}.}
\section{Concluding remarks}
\label{sec:conclusion}
\new{We introduced the concept of Leakage Templates to abstractly describe specific \new{side channels}, and determine relations between input parameters that when satisfied can trigger specific microarchitectural behavior.} \LT{}s allow to \new{automate} identifying code snippets that are vulnerable to \new{side-channel} attacks in application binaries for certain target architectures. As such, they enable attackers and analysts to identify potential \new{side channels} in applications.

As details on microarchitectural aspects such as cache eviction policy, prefetching and previction are scarce, derivation of \LT{}s is challenging. Expressive specifications for testcases are required, a large set of 
inputs has to be explored, and low noise measurement setups are essential. Also, techniques to automate discovery of relations between code, data, and leakage behavior are needed.

To address those challenges, we proposed \framework, which leverages \textit{instruction fuzzing}, \new{\textit{instructions' operand mutation}} and \textit{statistical analysis} to explore underspecified behavior of microarchitectural optimisations. \new{\framework{'s} high-level goal is to \textit{facilitate} the understanding of microarchitectural behavior. Therefore, we expect that the user has some prior knowledge regarding the existence of undocumented behavior or potential information leakage.} 

We showed the utility of templates produced by \framework by identifying five novel side-channel attack primitives (for an ARM Cortex-A53 core) and used \framework{} to reverse engineer the A53's branch predictor \new{structure}. 
\new{Also, we showed how LT{s} can be used to identify an instance of the specified side channel in a target binary.}
We plan to extend \framework further, e.g., to reverse engineer port contention behavior by setting the measured state to be the execution time of various types of instructions. A generated request would contain a number of instructions of the same type. Then, the analysis would only compare execution times and return the relation 
between executed and measured instructions. One can also use execution time measurements to extract the operand/runtime correlation of a multi-clock-cycle multiplier.

\section*{Acknowledgement}
This work was supported in part by the German Federal Ministry of Education and Research (BMBF) through funding for the CISPA-Stanford Center for Cybersecurity (FKZ: 13N1S0762). 

\bibliographystyle{plain}
\balance
\bibliography{previction.bib}

 \appendix
\section{Asmregex pattern for the prefech leakage template}
\label{sec:asmregex_pattern}

\begin{figure}
   \includegraphics[width=0.65\linewidth]{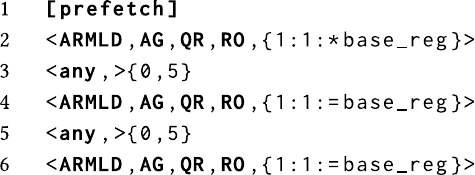}
    \caption{\new{An \emph{asmregex} pattern to identify instances of the prefetching \LT{}}}
    \label{fig:asmregex_pattern_prf}
    \Description{Asmregex pattern code listing.}
\end{figure}

\autoref{fig:asmregex_pattern_prf} shows the asmregex pattern that is used to identify instances of the prefetching \LT{} in binaries. It specifies a sequence of three load instructions with a maximum distance of 5 other instructions between them.

Each of the load instructions (\texttt{ARMLD}) is further required to fulfill the following properties:
\begin{itemize}
    \item Operand 0 (\texttt{AG}) is a general purpose register (e.g., \texttt{x0}).
    \item Operand 1 (\texttt{QR}) is a square bracket, followed by a general purpose register (e.g., \texttt{[x1})
    \item Operand 2 (\texttt{RO}) is a general purpose register, optionally followed by a closing square bracket (e.g. \texttt{x2} or \texttt{x2]})
\end{itemize}
Therefore, this pattern line matches both of the following instructions:
\begin{itemize} 
    \item \texttt{ldr x0, [x1, x2]}
    \item \texttt{ldr x0, [x1, x2, lsl\#3]}
\end{itemize}
In addition, all three load instruction pattern lines contain a backreference to constrain operand 1. In particular, the pattern requires that operand 1 is the same for all three load instructions. The backreference is initialized in line 2 and is used as a constraint in lines 4 and 6. This constraint is added as a heuristic to ensure that all three loads load from addresses that are close to each other, which is a mandatory requirement to trigger prefetching.

\lstdefinelanguage{asmregex}{
    keywords = {ARMLD,AG,QR,RO,any},
    morecomment=[l]{\#},
    morecomment=[s][keywordstyle]{[}{]}
}

 \section{Reverse Engineering Cache Slice Selection}
\label{sec:discussion:slice}

To improve the performance of the last-level cache, modern processors divide it into multiple slices that can be accessed in parallel by different cores.
Intel CPUs use a deterministic hash function to select cache slices.
A reverse engineering approach for cache slice selection of Intel CPUs
based on Prime+Probe is suggested in~\cite{DBLP:journals/iacr/ApececheaES15}.

The task is done in three steps: (1) $m$ data blocks residing in the same slice are identified; (2) equations for slice mapping are generated; (3) the used hash function is recovered. 
In what follows we show how these steps can be done using \framework.
Since this task is similar to the eviction example from \autoref{sec:eviction}, the classifier classifies the output based on the existence of data block(s) in memory. This can be achieved through probing and timing, or by simply inspecting the content of the cache.

\paragraph{Step 1} An address is pre-loaded followed by $b$ loads to the same cache set causing eviction of this line. Then one of the loads is removed and the execution is repeated. If eviction no longer occurs, it indicates that the removed load address maps to the same slice as the pre-loaded address. 
This step is repeated until $m$ data blocks are identified.
This experiment can be done by running the \requestShort:

$\rep{\pre{\memtwo{\tagone}{\setone} \s(\memtwo{\tagx}{\setone}\s \cdots \s\memtwo{\tagxb}{\setone})}\memtwo{\tagone}{\setone}}{1000}$\\
where each experiment is repeated $1000$ times. 

\paragraph{Step 2} In this step more data blocks that reside in the same slice are generated. This is done similar to step one. The only difference is that the $m$ generated data blocks are used to prime the cache (instead of one block). This step is repeated until a large number of memory blocks that map to the same slice is identified.
This experiment is represented by the following \requestShort:

$\rep{\pre{\memtwo{\tagone}{\setone} \; \cdots \; \memtwo{\tagm}{\setone} \s(\memtwo{\tagx}{\setone} \; \cdots \; \memtwo{\tagxb}{\setone})}\memtwo{\tagone}{\setone} \; \cdots \; \memtwo{\tagm}{\setone}}{1000}$\\
where each experiment is repeated $1000$ times.
The found blocks are used to generate a system of equations/matrix. This can be implemented within \framework's analyzer function.

\paragraph{Step 3} Based on the generated system of equations, the hash function for the slice mapping can be recovered. This is done by modelling this function as a concatenation of binary linear functions. These functions are then determined based on the matrix representing the system equations.
This step can also be implemented in \framework's analyzer function, given the individual bits of each physical address provided by the Classifier.

\end{document}